\documentclass[sigconf]{acmart}
\usepackage[utf8]{inputenc}
\usepackage[T1]{fontenc}
\usepackage{hyphenat}
\usepackage{xspace}
\usepackage{amsmath}
\usepackage{amsfonts}
\usepackage{hyperref}
\usepackage{url}
\usepackage{booktabs}
\usepackage{multirow}
\usepackage{subfig}
\usepackage{makecell}
\usepackage{caption}
\usepackage{minibox}
\usepackage{bbm}
\usepackage{graphicx}
\usepackage{balance}
\usepackage{mathtools}
\usepackage{color}
\usepackage{marvosym}
\usepackage{ifthen}
\usepackage{textcomp}
\usepackage{enumitem}
\usepackage{verbatim}
\usepackage{algorithm}
\usepackage{algorithmic}
\usepackage{numprint}

\usepackage{amsthm}
\theoremstyle{plain}

\definecolor{MyRed}{rgb}{0.6,0.0,0.0} 
\definecolor{MyBlack}{rgb}{0.1,0.1,0.1} 
\newcommand{\inred}[1]{{\color{MyRed}\sf\textbf{\textsc{#1}}}}

\newcommand{\frameit}[2]{
  \begin{center}
  {\color{MyRed}
  \framebox[.9\columnwidth][l]{
    \begin{minipage}{.85\columnwidth}
    \inred{#1}: {\sf\color{MyBlack}#2}
    \end{minipage}
  }\\
  }
  \end{center}
}

\newcommand{\chatoDisplayMode}[1]{#1}

\newcommand{\note}[2][]{\chatoDisplayMode{\def\@tmpsig{#1}\frameit{{\Pointinghand} Note}{#2\ifx \@tmpsig \@empty \else \mbox{ --\em #1}\fi}}}
\newcommand{\todo}[2][]{\chatoDisplayMode{\def\@tmpsig{#1}\frameit{{\Writinghand} To-do}{#2\ifx \@tmpsig \@empty \else \mbox{ --\em #1}\fi}}}

\newcommand{\abbrevStyle}[1]{#1}

\newcommand{\ie}{\abbrevStyle{i.e.}\xspace}
\newcommand{\eg}{\abbrevStyle{e.g.}\xspace}
\newcommand{\cf}{\abbrevStyle{cf.}\xspace}

\newcommand{\vs}{\abbrevStyle{vs.}\xspace}
\newcommand{\etc}{\abbrevStyle{etc.}\xspace}

\newcommand{\Secref}[1]{Sec.~\ref{#1}}
\newcommand{\Eqnref}[1]{Eq.~\ref{#1}}

\newcommand{\Figref}[1]{Fig.~\ref{#1}}

\newcommand{\xhdr}[1]{\vspace{1.7mm}\noindent{{\bf #1.}}}

\newcommand{\xhdrNoPeriod}[1]{\vspace{1.7mm}\noindent{{\bf #1}}}

\newcommand{\textcite}[1]{\citeauthor{#1} \shortcite{#1}}

\newcommand{\hide}[1]{}

\newcommand{\affilSize}{9pt}
\newcommand{\authorbox}[3]{
  \minibox[c]{
    #1\\
    {\fontsize{\affilSize}{\affilSize}\selectfont{}#2}\\
    {\fontsize{\affilSize}{\affilSize}\selectfont{}#3}
  }
}

\newcommand\blfootnote[1]{%
  \begingroup
  \renewcommand\thefootnote{}\footnote{#1}%
  \addtocounter{footnote}{-1}%
  \endgroup
}

\makeatletter
\newcommand{\iffont}[2]{\ifthenelse{\equal{\f@family}{#1}}{#2}{}}
\makeatother

\iffont{ptm}{
  \usepackage{mathptmx}

  \DeclareSymbolFont{greek}{OML}{cmm}{m}{n}
  \DeclareMathSymbol{\alpha}{\mathalpha}{greek}{"0B}
  \DeclareMathSymbol{\beta}{\mathalpha}{greek}{"0C}
  \DeclareMathSymbol{\gamma}{\mathalpha}{greek}{"0D}
  \DeclareMathSymbol{\delta}{\mathalpha}{greek}{"0E}
  \DeclareMathSymbol{\epsilon}{\mathalpha}{greek}{"0F}
  \DeclareMathSymbol{\zeta}{\mathalpha}{greek}{"10}
  \DeclareMathSymbol{\eta}{\mathalpha}{greek}{"11}
  \DeclareMathSymbol{\theta}{\mathalpha}{greek}{"12}
  \DeclareMathSymbol{\iota}{\mathalpha}{greek}{"13}
  \DeclareMathSymbol{\kappa}{\mathalpha}{greek}{"14}
  \DeclareMathSymbol{\lambda}{\mathalpha}{greek}{"15}
  \DeclareMathSymbol{\mu}{\mathalpha}{greek}{"16}
  \DeclareMathSymbol{\nu}{\mathalpha}{greek}{"17}
  \DeclareMathSymbol{\xi}{\mathalpha}{greek}{"18}
  \DeclareMathSymbol{\pi}{\mathalpha}{greek}{"19}
  \DeclareMathSymbol{\rho}{\mathalpha}{greek}{"1A}
  \DeclareMathSymbol{\sigma}{\mathalpha}{greek}{"1B}
  \DeclareMathSymbol{\tau}{\mathalpha}{greek}{"1C}
  \DeclareMathSymbol{\upsilon}{\mathalpha}{greek}{"1D}
  \DeclareMathSymbol{\phi}{\mathalpha}{greek}{"1E}
  \DeclareMathSymbol{\chi}{\mathalpha}{greek}{"1F}
  \DeclareMathSymbol{\psi}{\mathalpha}{greek}{"20}
  \DeclareMathSymbol{\omega}{\mathalpha}{greek}{"21}
  \DeclareMathSymbol{\varepsilon}{\mathalpha}{greek}{"22}
  \DeclareMathSymbol{\vartheta}{\mathalpha}{greek}{"23}
  \DeclareMathSymbol{\varpi}{\mathalpha}{greek}{"24}
  \DeclareMathSymbol{\varrho}{\mathalpha}{greek}{"25}
  \DeclareMathSymbol{\varsigma}{\mathalpha}{greek}{"26}
  \DeclareMathSymbol{\varphi}{\mathalpha}{greek}{"27}
  \DeclareSymbolFont{otone}{OT1}{cmr}{m}{n}
  \DeclareMathSymbol{\Gamma}{\mathalpha}{otone}{0}
  \DeclareMathSymbol{\Delta}{\mathalpha}{otone}{1}
  \DeclareMathSymbol{\Theta}{\mathalpha}{otone}{2}
  \DeclareMathSymbol{\Lambda}{\mathalpha}{otone}{3}
  \DeclareMathSymbol{\Xi}{\mathalpha}{otone}{4}
  \DeclareMathSymbol{\Pi}{\mathalpha}{otone}{5}
  \DeclareMathSymbol{\Sigma}{\mathalpha}{otone}{6}
  \DeclareMathSymbol{\Upsilon}{\mathalpha}{otone}{7}
  \DeclareMathSymbol{\Phi}{\mathalpha}{otone}{8}
  \DeclareMathSymbol{\Psi}{\mathalpha}{otone}{9}
  \DeclareMathSymbol{\Omega}{\mathalpha}{otone}{10}
  \DeclareSymbolFont{syms}{OML}{cmm}{m}{it}
  \DeclareMathSymbol{\partial}{\mathord}{syms}{"40}
  \DeclareMathAlphabet{\mathbold}{OML}{cmm}{b}{it}
  \DeclareSymbolFont{largesymbols}{OMX}{cmex}{m}{n}

}

\newcommand{\B}{Broccoli\xspace}
\newcommand{\GitHubURL}{\url{https://github.com/epfl-dlab/broccoli}}

\copyrightyear{2020}
\acmYear{2020}
\setcopyright{iw3c2w3}
\acmConference[WWW '20]{Proceedings of The Web Conference 2020}{April 20--24, 2020}{Taipei, Taiwan}
\acmBooktitle{Proceedings of The Web Conference 2020 (WWW '20), April 20--24, 2020, Taipei, Taiwan}
\acmPrice{}
\acmDOI{10.1145/3366423.3380209}
\acmISBN{978-1-4503-7023-3/20/04}

\title{Broccoli: Sprinkling Lightweight Vocabulary Learning into Everyday Information Diets}

\begin{document}

\author{
\authorbox{Roland Aydin*}{Helmholtz-Zentrum Geesthacht}{roland.aydin@hzg.de}\hspace{7mm}
\authorbox{Lars Klein*}{EPFL}{lars.klein@epf\/l.ch}\hspace{9mm}
\authorbox{Arnaud Miribel\textsuperscript{\textdagger}}{byrd valley}{arnaudm@byrdvalley.com}\hspace{9mm}
\authorbox{Robert West}{EPFL}{robert.west@epf\/l.ch}
}

\renewcommand{\shortauthors}{Roland Aydin, Lars Klein, Arnaud Miribel, and Robert West}


\begin{abstract}
The learning of a new language remains to this date a cognitive task that requires considerable diligence and willpower, recent advances and tools notwithstanding. In this paper, we propose \emph{Broccoli,} a new paradigm aimed at reducing the required effort by seamlessly embedding vocabulary learning into users' everyday information diets. This is achieved by inconspicuously switching chosen words encountered by the user for their translation in the target language. Thus, by seeing words in context, the user can assimilate new vocabulary without much conscious effort. We validate our approach in a careful user study, finding that the efficacy of the lightweight Broccoli approach is competitive with traditional, memorization-based vocabulary learning. The low cognitive overhead is manifested in a pronounced decrease in learners' usage of mnemonic learning strategies, as compared to traditional learning. Finally, we establish that language patterns in typical information diets are compatible with spaced-repetition strategies, thus enabling an efficient use of the Broccoli paradigm. Overall, our work establishes the feasibility of a novel and powerful ``install-and-forget'' approach for embedded language acquisition.
\end{abstract}

\maketitle

{\fontsize{8pt}{8pt} \selectfont
\textbf{ACM Reference Format:}\\
Roland Aydin, Lars Klein, Arnaud Miribel, and Robert West.
2020.
Broccoli: Sprinkling Lightweight Vocabulary Learning into Everyday Information Diets.
In
\textit{Proceedings of The Web Conference 2020 (WWW '20), April 20--24, 2020, Taipei, Taiwan.}
ACM, New York, NY, USA, 11 pages. \url{https://doi.org/10.1145/3366423.3380209}
}

\blfootnote{*Equal contribution.
\hspace{1mm}
\textsuperscript{\textdagger}Work performed at EPFL.}

\section{Introduction}
\label{sec:Introduction}

Most people would, all else being equal, prefer to know more foreign languages than they do.
But mustering the willpower necessary to establish a habit of learning a foreign language presents a high barrier.
For instance, every time a learner studies with the help of classic vocabulary training software, this will have required a conscious, disciplined decision to do so. In order to alleviate the often tedious vocabulary learning process, current approaches typically either try to improve the experience (\eg, via gamification) or at least to minimize the necessary time by maximizing the efficiency of language acquisition (\eg, spaced repetition via flashcards). Despite these improvements, there remain massive barriers that prevent many people from acquiring the languages they would ideally like to know.

\xhdr{\B}
To lower these barriers, we propose a novel vocabulary training paradigm, \textit{\B,} which aims to embed vocabulary learning into everyday information diets, such as Web browsing or e-book reading, with a minimal cognitive footprint.
\B aims not only to integrate into a user's daily information diet, but rather to properly hide the tedious process of language acquisition in it. 
To do so, \B augments text that the user is reading of her own accord with passive exposures to vocabulary from a foreign target language. It chooses words from the text being read, translates them in context, and re-inserts the translation into the original text. Our tool is intended as a constant companion, removing the user's responsibility to make an active decision to revise vocabulary. As such, \B offers an ``install-and-forget'' approach to vocabulary learning. Once installed, it will steadily enhance the user's vocabulary, minimizing the conscious effort on her part.
With hours dedicated to browsing the Web each day, even a slow but steady and sustainable learning progression---barely noticed by the user---has the potential to result in significant new language skills.

To make the process of learning vocabulary by passive exposure as efficient as possible, we introduce a processing pipeline for scoring and prioritizing words. It combines concepts from spaced repetition for optimizing review times, a language model for estimating the guessability of words in context, and machine translation software.

\xhdr{Results}
Based on an implemented prototype (available as an open-source browser plugin), we conduct a carefully controlled within-subject user study, where we validate the effectiveness of the \B paradigm vis-\`a-vis a conventional vocabulary learning approach in which users explicitly memorize word translations shown to them in table form.
\B-assisted learners achieve surprisingly strong results, obtaining short-term word retention rates 50\% higher than in the table-based approach, and long-term retention rates indistinguishable from those in the table-based approach.
Moreover, \B's low cognitive overhead is manifested in a much decreased usage of artificial mnemonic learning strategies as compared to traditional learning.

Finally, we establish that language patterns in typical information diets (Web browsing and e-book reading) are compatible with spaced  repetition, thus enabling an efficient use of the \B paradigm.
While our user study focuses on Wikipedia, the compatibility analysis additionally investigates fiction books.

Taken together, these results provide strong arguments in favor of the paradigm we have developed.

\xhdr{Contributions}
This paper makes the following core contributions:
\begin{itemize}
    \item We propose \B, a novel paradigm for vocabulary learning sprinkled into everyday information diets (\Secref{sec:Broccoli paradigm}).
    \item We implement a prototype as a browser plugin (\Secref{sec:Implemented prototype}).
    \item In a controlled user study (\Secref{sec:Experimental design}), we validate \B's effectiveness with strong results (\Secref{sec:Results}).
    \item We establish the compatibility of \B with different online information diets (\Secref{sec:Compatibility}).
\end{itemize}

We begin by introducing related work (\Secref{sec:Related work}) and conclude with a discussion of implications, limitations, and future work (\Secref{sec:Discussion}).


\section{Related work}
\label{sec:Related work}

\xhdr{Learning from context}
The power of using contextual clues for learning has long been established and is for the most part well understood. ``Incidental learning from context'' may in fact be the natural mode of learning \cite{nagy1985learning}. Indeed, most language acquisition seems to rely on contextual clues \cite{sternberg1987most}, rather than memorization as in the flashcard approach, and learning from context is an effective learning strategy \cite{carnine1984utilization} that also works for normal reading \cite{nagy1987learning}. The power of contextual information for learning seems to be language\hyp independent to some degree, \eg, also applying to Japanese students learning English, whose performance was shown to benefit not only from repetitions but also from contextual information \cite{webb2008effects}. 

The degree to which context captures information about a word is varied \cite{beck1983vocabulary}, which supports the usage of language models to further classify the degree to which a specific instance of context conveys information about a word. Related, learning in context also enhances word retrieval \cite{van2018contextual}. This mode of learning in itself seems to be improvable over time simply through usage (as with Broccoli), as established by a meta study investigating approaches aimed at learning words from context which concludes that practice of learning from context was more important than the specific strategy used \cite{kuhn1998teaching}.

\xhdr{Spaced repetition}
The deep scientific literature on spaced repetition has long established that \textit{``repetitio est mater studiorum''} \cite{dempster1989spacing,melton1970situation,godwin2010emerging, ausubel1965effect}, and that the spacing of repetition intervals can be adapted to better fit human memory retention behavior, also via machine learning tools \cite{tabibian2019enhancing,settles2016trainable,atkinson1972optimizing}. This approach has found widespread practical use in the form of flashcard applications. Among those, SuperMemo pioneered research on memory models and spaced repetition \cite{wozniak2005two, wozniak2007supermemo}, which later culminated in Anki \cite{Anki} (an open source spinoff) and Mnemosyne~\cite{WelcomeMnemosyneProject}, among others. 

Although there is a multitude of language learning tools available both on- and offline, such applications for language acquisition, while highly effective, do not mesh easily with a user's daily habits. They for the most part only superficially leverage the daily hours perusing Web-based content and require conscious effort and discipline even when automatically scheduled. 

Even for learning tools with lower cognitive footprints, there is to the best of our knowledge no scientifically validated model which leverages passive learning in the context of an existing everyday information diet.


\section{The Broccoli paradigm}
\label{sec:Broccoli paradigm}

In classic, memorization\hyp based vocabulary training software tools, such as those based on virtual flashcards, the core problem consists in deciding which word to teach or revise next, from a base pool of words that are to be ultimately acquired by the user.
This task can be understood as an optimization problem with the goal of maximizing the long-term retention of as many words as possible with as few repetitions as possible.
This alone is a hard problem, and the literature on it is vast (\cf \Secref{sec:Related work}).

The system developed here, \B, is expected to optimize the same objective, but under added constraints:
when choosing what words to revise next, \B
\begin{itemize}
    \item must pick only words from the text that the user is currently reading;
    \item should pick words whose meaning can easily be guessed from their current context;
    \item must translate the chosen words in their context and embed the appropriate translations into the present text.
\end{itemize}

In order to deal with all of these constraints in a manageable way, \B was conceived as a pipeline architecture, a common design pattern in natural language processing \cite{manning-EtAl:2014:P14-5}.
Given the text the user is currently reading (of her own accord) and the target language the user intends to learn, \B needs to choose some words and replace them with their appropriate translations in the target language.
The base set of candidates for translation is therefore obtained by tokenizing and lemmatizing%
\footnote{We generally operate on lemmas rather than inflected forms and shall use the terms \textit{lemma} and \textit{word} interchangeably.}
the currently read text.
The base set is then iteratively scored and filtered in four stages:
\begin{enumerate}
    \item \textbf{Tutor\hyp based word scoring (\Secref{sec:Word scheduling}).} Determine how much each word would benefit from being revised right now.
    \item \textbf{Context\hyp based word scoring (\Secref{sec:Context-based word scoring}).} Determine in how typical a context each word appears in the current text.
    \item \textbf{Word set selection (\Secref{sec:Word set selection}).} Based on the scores from stages 1 and 2 and a user-defined translation density parameter, select a set of words to translate.
    \item \textbf{Word translation (\Secref{sec:Word translation}).} Machine\hyp translate each selected word in its present context into the target language.
\end{enumerate}

In the next section, we will present in detail how we implemented each pipeline stage.

We emphasize that the main goal of this paper is to establish the feasibility of \B's philosophy of sprinkling lightweight vocabulary learning into everyday information diets, so our focus is on building a simple, modular prototype and evaluating it with real users.


\section{Implemented prototype}
\label{sec:Implemented prototype}
We have designed a concrete realization of the Broccoli paradigm as a browser plugin, which we make available as open-source software.%
\footnote{\GitHubURL}
The client frontend is decoupled from the Broccoli logic and communicates with a server\hyp hosted backend implementation of the pipeline.
This way, our API could easily be adapted to situations beyond Web browsing, such as e-books or newsreaders.

In our prototype, the first two pipeline stages provide probabilities. They can be merged in a meaningful way to estimate the probability of the user understanding a given word. From this probability, together with metrics provided by the tutor algorithm, we derive an overall word score. This score is the basis for decisions made during word set selection. Finally, we machine\hyp translate the selected words.

\subsection{Tutor-based word scoring}
\label{sec:Word scheduling}

The purpose of the tutor algorithm is to score each lemma with respect to how much it would benefit from being revised (\ie, shown as a translation) right now.
The central quantity here is a lemma's \textbf{recall probability} $R$, the probability that the user can recall the original meaning of a translated lemma from her memory.
To model $R$, we closely follow the vetted SuperMemo model \cite{wozniak2005two}, where $R$ decays exponentially as a function of the time $t$ since the word was last successfully revised:
\begin{equation}
\label{eq:R}
    R = 2^{-t/H},
\end{equation}
where $H$ is the lemma's \textbf{half-life,} which captures the lemma's stability in the user's memory.

The model further assumes that a lemma benefits in two ways from being successfully revised:
first, its recall probability is reset to $R=1$ (and $t$ to 0), and second, the half-life $H$ is multiplied by a \textbf{boosting factor} $\gamma$:
\begin{equation}
\label{eq:H}
    H \leftarrow \gamma \, H.\\
\end{equation}
The boosting factor $\gamma$ is in turn modeled as a decreasing function of the current $H$ and $R$ (parametrized by constants $a,b,c,d > 0$, based on SuperMemo's empirically fitted values \cite{wozniak2005two}):
\begin{equation}
\label{eq:gamma}
\gamma = a \, H^{-b} c^{-R} + d.
\end{equation}

In conventional spaced\hyp repetition paradigms (such as those based on flashcards), it is explicitly observed whether a revision was successful or not (the user answers either correctly or incorrectly), whereas in \B this is not the case: we do not know if the user understood a translation in context.
To circumvent this issue, we may simply model every exposure as successful.
Although this is not true, it is not a problem in practice because all lemmas are affected equally by this simplification.%
\footnote{
\B can also be modified to incorporate explicit feedback, \cf\ \Secref{sec:Discussion}.
}

Later in the pipeline (\Secref{sec:Word set selection}), $R$ and $\gamma$ are combined with the context-based guessability of the lemma (\Secref{sec:Context-based word scoring}) to obtain its final score used for selecting words to translate.

\subsection{Context-based word scoring}
\label{sec:Context-based word scoring}

The core concept of Broccoli is the idea of learning vocabulary via passive exposure.
We are thus primarily interested in word occurrences whose meaning is evident from their context, as this will presumably increase the learning effect and will disrupt the user's natural reading less.
To approximate the ease with which a word's meaning can be inferred from its present context by a human reader, we employ a powerful neural\hyp network\hyp based language model, AWD-LSTM \cite{merity2017regularizing,FastAIAWDLSTM}, trained on the WikiText-103 corpus~\cite{merityPointerSentinelMixture2016}.
The model scans text from left to right and, at any point, outputs a probability distribution over all possible next words.

Running the pretrained model on the text currently being read by the user, we score each word with the probability attributed to it by the language model based on the word's left context and refer to this score as the word's \textbf{guessability} $G$.
In the next pipeline stage, $G$ is combined with the predictions of the tutor algorithm in order to obtain the word's final fitness for selection.

\subsection{Word set selection}
\label{sec:Word set selection}

For simplicity, we assume that recalling a word from memory and understanding it from context are two independent events. With $R$ the probability of recalling a word from memory (\Eqnref{eq:R}) and $G$ the probability of guessing the same word from context (\Secref{sec:Context-based word scoring}),
$P = R + G - RG$
describes the probability of understanding the word at the given time in the given context.

Since we are interested in words that (1)~are easy to understand and (2)~would experience a large boost $\gamma$ in half-life (\Eqnref{eq:H}), we combine $P$ and $\gamma$ into a single priority score by multiplying them.
Finally, we greedily select the $k$ highest\hyp priority words (where $k$ is determined based on a user\hyp specified translation density preference) and pass them to the next pipeline stage.

\subsection{Word translation}
\label{sec:Word translation}

In the final pipeline stage, for each selected word occurrence $w$, we translate $w$'s entire sentence using the Microsoft Translator Text API \cite{TranslatorTextAPI}.
The API provides an alignment matrix that allows us to recover $w$'s correct context-based translation in the target language.
By translating words in context, we avoid errors due to homonymy.

\begin{figure*}[t]   
\centering
\subfloat[Table-based learning section]{\includegraphics[width=0.25\linewidth, keepaspectratio,trim=10 0 10 80,clip]{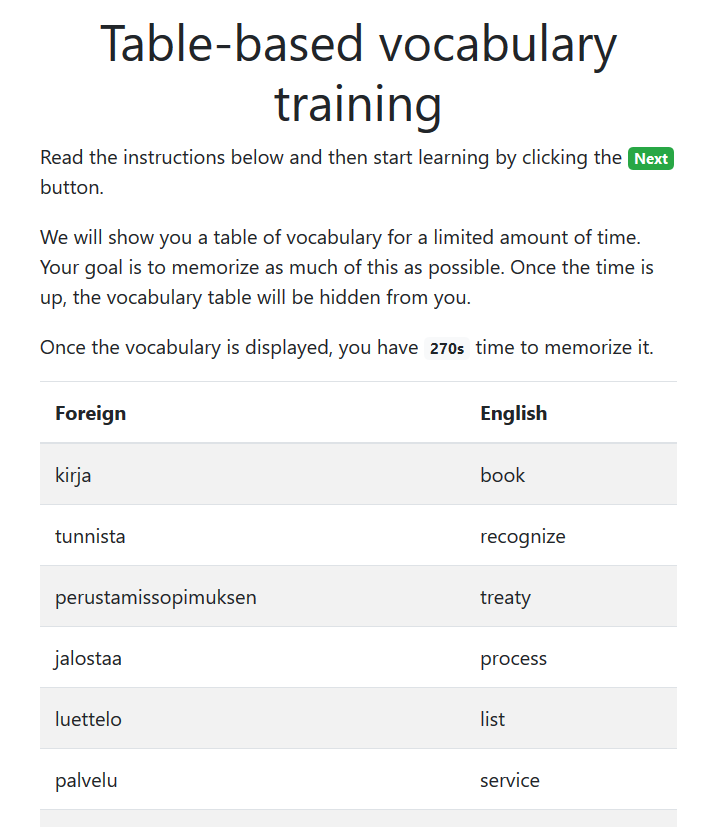}\label{fig:screenshot_table}}
\subfloat[Broccoli-based learning section]{\includegraphics[width=0.25\linewidth, keepaspectratio,trim=0 0 0 0,clip]{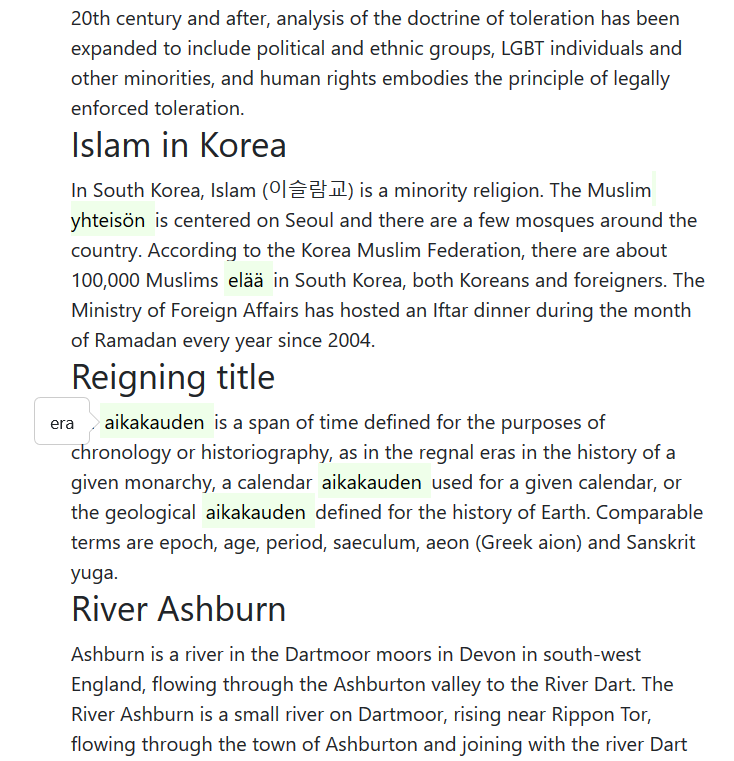}\label{fig:screenshot_wiki_closeup}}
\subfloat[Multiple-choice testing section]{\includegraphics[width=0.25\linewidth, keepaspectratio,trim=0 0 0 80,clip]{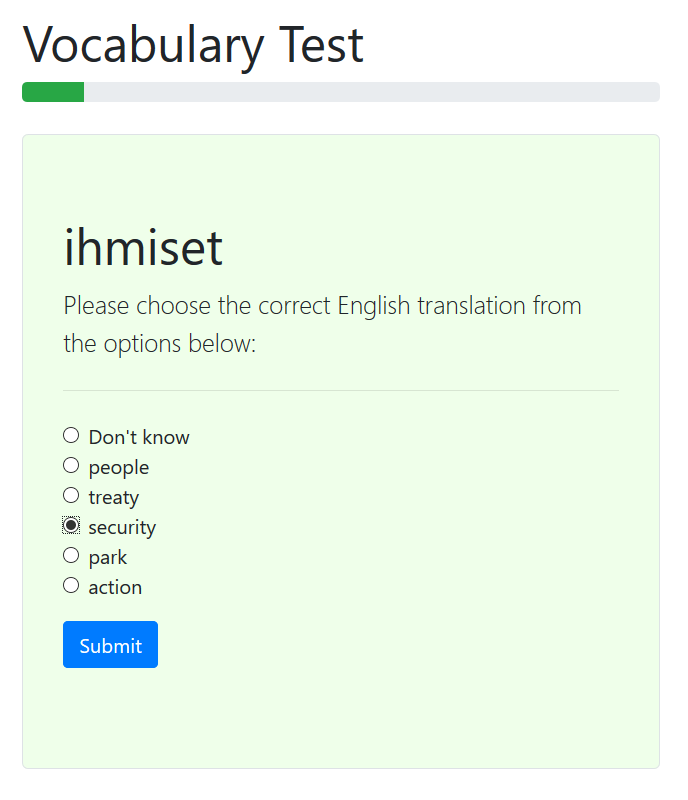}\label{fig:screenshot_MC}}
\subfloat[Fill-the-gap testing section]{\includegraphics[width=0.25\linewidth, keepaspectratio,trim=0 0 0 80,clip]{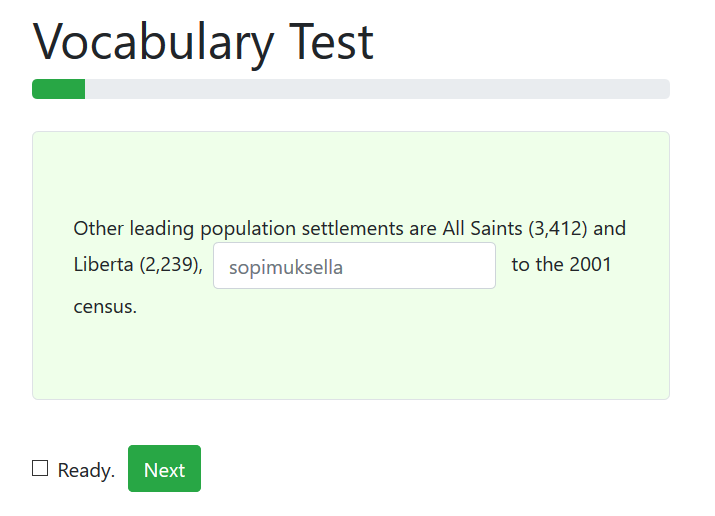}\label{fig:screenshot_FTG}} 
\caption{
Screenshots of user interface from user study:
(a--b) Learning sections.
(c--d) Testing sections.
In \B-based learning~(b), Finnish words inserted into English text are highlighted in green; clicking Finnish word (\eg, \textit{aikakauden}) reveals English translation (\eg, \textit{era}).
}
\label{fig:screenshots}
\end{figure*}

In our prototype, translated tokens are marked with a subtly colored background, and by clicking on them, the user can reveal the original tokens (\Figref{fig:screenshot_wiki_closeup}), although these features are not strictly necessary for \B.


\begin{figure}[t]
    \resizebox{0.6\linewidth}{!}{
        \includegraphics{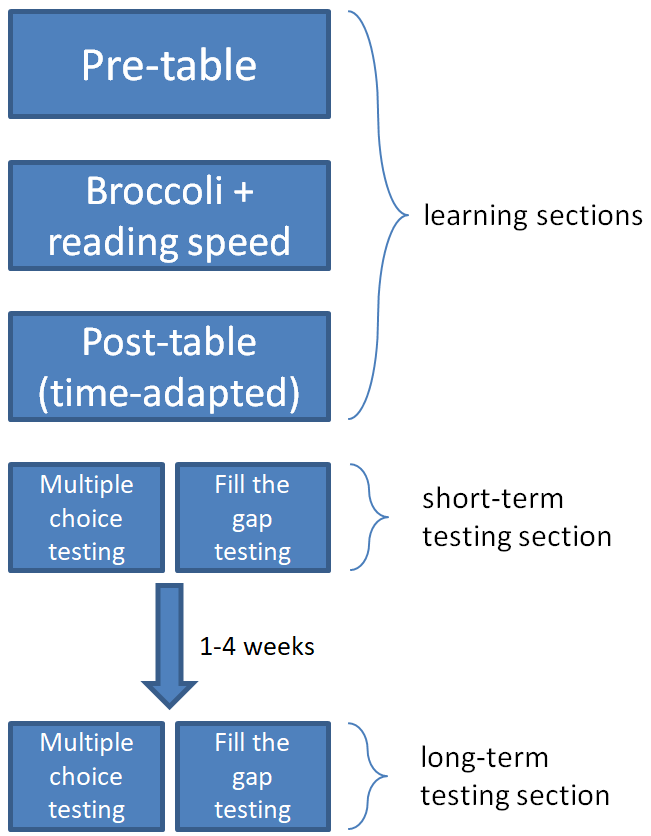}
    }
    \caption{Experimental design of the user study (\Secref{sec:Experimental design}).}
    \label{fig:study_structure}
\end{figure}  

\section{User study: Experimental design}
\label{sec:User_study}
\label{sec:Experimental design}

The primary research question to be answered by the user study is,
\textit{How does Broccoli compare to traditional vocabulary acquisition?}
In this comparative approach, the baseline to which we compare Broccoli is explicit memorization of translation pairs presented in a table (\Figref{fig:screenshot_table}).
In total, we answer five research questions:

\begin{enumerate}
    \item[\textbf{RQ1}] How does Broccoli compare to traditional, memorization\hyp based vocabulary acquisition? (\Secref{sec:Results RQ1})
    \item[\textbf{RQ2}] Does revealing explicit translations during Broccoli\hyp enriched reading improve word retention? (\Secref{sec:Results RQ2})
    \item[\textbf{RQ3}] By how much is reading slowed down when using Broccoli? (\Secref{sec:Results RQ3})
    \item[\textbf{RQ4}] What are the benefits of using a language model for selecting words to translate? (\Secref{sec:Results RQ4})
    \item[\textbf{RQ5}] How effectively does Broccoli reduce the cognitive load (as manifested in the usage of unnatural mnemonic tricks) during learning? (\Secref{sec:Results RQ5})
\end{enumerate}

Note that the study does not attempt to validate the effectiveness of the tutor algorithm (\Secref{sec:Word scheduling}), which closely follows the SuperMemo model \cite{wozniak2005two} and is hence orthogonal to the core idea of embedding vocabulary learning into everyday information diets.

\xhdr{Overall design: within-subject study}
The above research questions are best addressed in a within-subject design, which circumvents many potential sources of variance inherent in between-subject designs.
In our setup, each participant learns vocabulary in 3 conditions:
\B, plus 2 table\hyp based conditions where the participant explicitly memorizes translation pairs (\Figref{fig:screenshot_table}; see below for the reasoning behind using 2 table-based conditions).

The overall structure of the study (\Figref{fig:study_structure}) comprises, per participant, 3 learning sections (one per condition) and 2 testing sections.
The learning sections and the first (short-term) testing section took place immediately after one another;
the second (long-term) testing section took place 1 to 4 weeks later.

The user study was implemented with English as the base language and Finnish as the target language. Finnish was chosen as a European language with few commonalities with other languages typically known by prospective participants.

We chose a common set of 78 English\slash Finnish word pairs that are used for all participants (details below).
Each participant learned each Finnish word in exactly one of the 3 conditions;
and each Finnish word was learned in each condition.

To achieve this design, we randomized the 78 words into 3 disjoint word groups.
All 26 words from the same word group were learned by the same participant in the same condition.
Since the 3 word groups can be distributed to the 3 conditions in 6 ways, participants were randomized into 6 disjoint participant groups.
This fully factorial, randomized design ensures that the 3 conditions can be compared in an unbiased way because each condition includes all participants and all words the same number of times.

\xhdr{Learning sections}
The 3 learning sections were arranged as follows:
first, a \textit{pre-table} section (``pre'' as it appears before the \B section);
second, a Broccoli section;
and third, a \textit{post-table} section (\Figref{fig:study_structure}).
For
kuvakaappauksia,
please refer to \Figref{fig:screenshots}a--b.

The pre-table was displayed for 4.5 minutes (\cf\ footnote~\ref{ftn:4.5min}), independent of the participant.

The \B section was split into 5--6 pages containing a total of 44--52 English text snippets from Wikipedia (details below).
A single page contained up to 10 snippets processed by \B.
Each of the 26 words appeared 7 times translated to Finnish; 13 words appeared only in ``easy'' English contexts as chosen by the language model (\Secref{sec:Context-based word scoring}), and 13 words only in randomly chosen English contexts, which allows us to evaluate the impact of using a language model.
Finnish words were marked by a light green background, and users could optionally click on these words to reveal the original English words (\Figref{fig:screenshot_wiki_closeup}).
No time limit was set for the \B section, but all user actions were recorded with timestamps.

We also included an additional page with 5 unaltered Wikipedia snippets, which allowed us to estimate the user's baseline reading speed and, by comparison to the time spent on the other pages, the time overhead $\Delta T$ incurred through \B.

Finally, the post-table was displayed for time $\Delta T$, thus ensuring that the user invested just as much time in memorizing words via the post-table as they had ``lost'' due to \B's translations.%
\footnote{\label{ftn:4.5min}The fixed time of 4.5 minutes used for the pre-table was determined in a pilot study in which we observed typical baseline reading speeds and \B overheads (about 10\%), which we then extrapolated to the length of the \B learning section.}

Placing \B between two table sections also avoids undue positional advantages for Broccoli which have been established for recall for both the earliest and the latest items \cite{murdock1962serial}.

\xhdr{Testing sections}
The testing sections consisted of a multiple-choice (MC, \Figref{fig:screenshot_MC}) and a fill-the-gap (FTG, \Figref{fig:screenshot_FTG}) component.
No time limits were imposed.
The MC component tested all 78 words in random order, in each step showing the user one Finnish word and 6 candidate answers:
``Don't know'' (always shown first) plus 5 candidate translations in English, randomly selected from the 78 words, one of them the correct translation.
The purpose of the FTG component was to test vocabulary retention in a context-based setting via free-form responses, thus avoiding exclusion-based reasoning as may be possible in the MC component.
As the FTG component was more time\hyp consuming than MC, we subsampled and tested only a third of the words in each condition.
The short-term and long-term tests were identical for a fixed user, with the exception that order randomization was performed independently.

\xhdr{Text and word selection for \B}
The texts users read in the \B condition were sampled from English Wikipedia, as a generally representative example of a typical reading diet. Wikipedia's common perception as both challenging and interesting were further reasons for this selection.
The study used introductory sections of Wikipedia articles (henceforth ``snippets''), as they are concise and self-contained.
Starting from 22,500 snippets (ranging between 100 and 2,500 tokens in length) as a base corpus, we performed lemmatization, selected the 500 most frequent English words (lemmas), and excluded stopwords, punctuation, proper nouns, words shorter than 3 characters, and words too similar to their analogues in Romance and Germanic languages (\eg, \textit{filosofi} = \textit{philosopher}).
This yielded 307 candidate words, from which we randomly selected the 78 words used in the study.
Finally, for each of the 3 groups of 26 words, we greedily selected 39 to 47 Wikipedia snippets to maximize coverage of the 26 words.

\xhdr{Participants}
We recruited 58 participants, mostly university students, at EPFL (Switzerland) and RWTH (Germany).
The exclusion criteria were prior knowledge in Finnish, insufficient (below B2) English language competency, and affiliation with our research groups.
The study was conducted at EPFL and RWTH in a supervised setting.
Participants received as remuneration the equivalent of US\$20.
They were 22.8 years old on average (SD 3.35) and 60\% male.
In terms of education, 9\% had a master's degree, 34\% a bachelor's degree, and 57\% a secondary school degree as the highest degree previously awarded.
The most prevalent mother tongues were
French (26\%),
German (16\%),
English (10\%), and
Italian (10\%).


\section{User study: Results}
\label{sec:Results}

Next, we present the results from the above\hyp described user study.
An anonymized version of the dataset is publicly available.%
\footnote{\GitHubURL}

\subsection{RQ1: \B\ \vs\ table\hyp based learning}
\label{sec:Results RQ1}

In both the multiple\hyp choice (MC) test and the fill-the-gap (FTG) test, there are three possible outcomes for each test word: either the user selected the \textit{correct} translation, the \textit{incorrect} translation, or no translation (\textit{``unknown''}).
To measure vocabulary retention in a single number for a given set of answers, we compute the fraction of correct answers and call this measure the \textbf{retention rate.}

\begin{figure}[tb]
\centering
\vspace{-3mm}
\subfloat[Multiple choice]{\includegraphics[width=\linewidth, keepaspectratio]{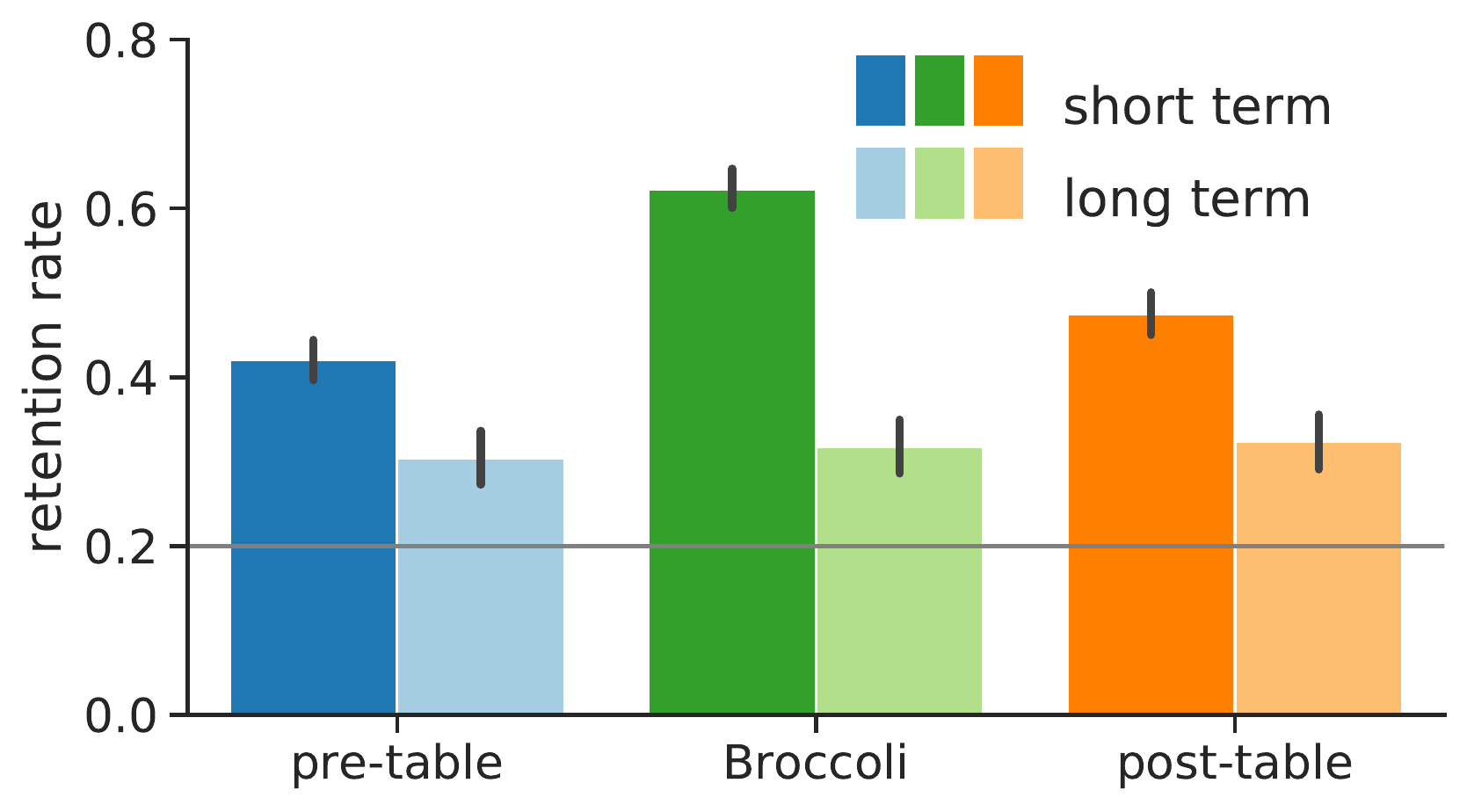}\label{fig:Broccoli_vs_Tables_MC}}

\subfloat[Fill the gap]{\includegraphics[width=\linewidth, keepaspectratio]{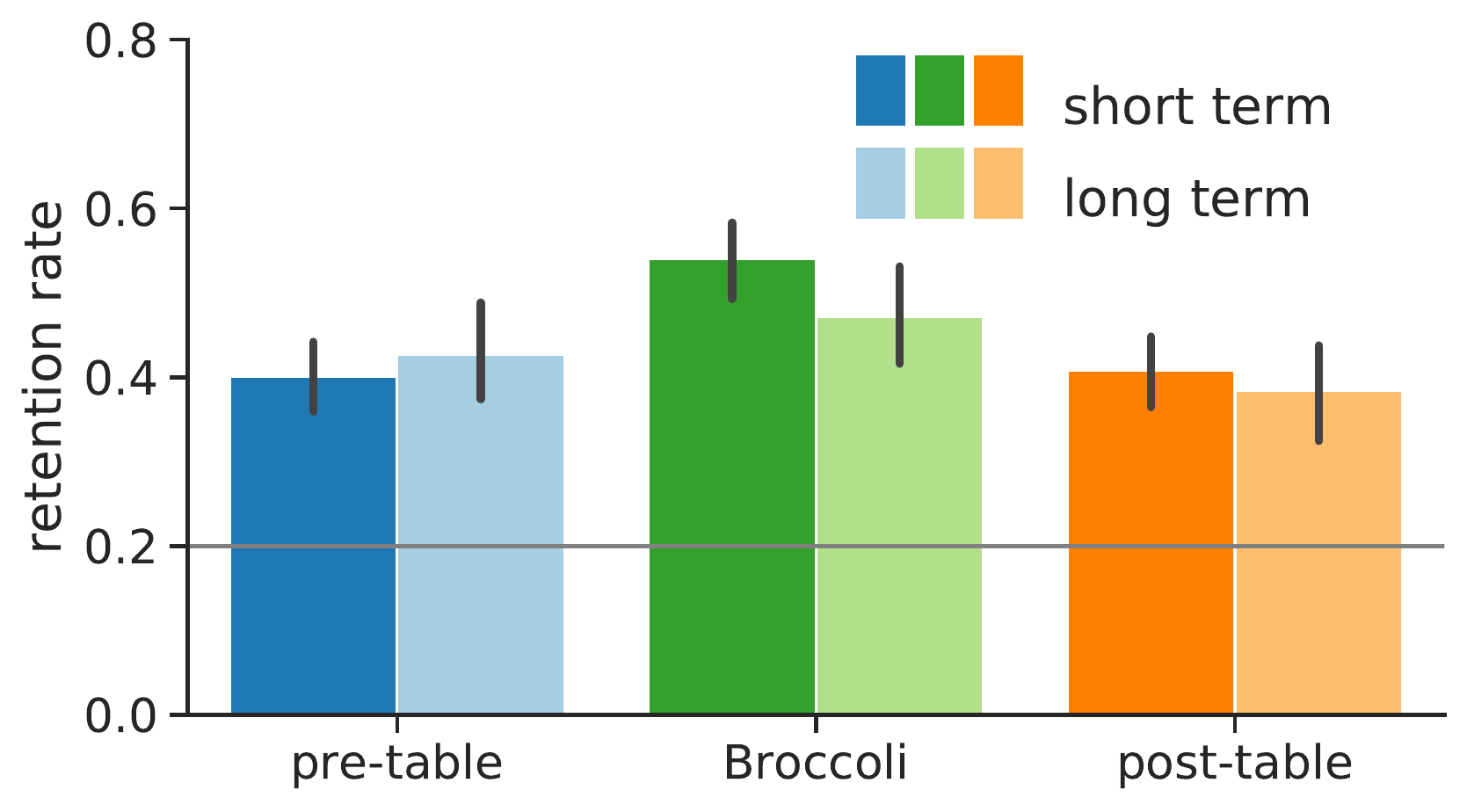}\label{fig:Broccoli_vs_Tables_FTG}}
    \caption{
    Broccoli \vs table-based learning:
    Retention rate (\Secref{sec:Results RQ1}) averaged over all users, with 95\% CIs; gray horizontal line: random guessing among 5 candidate translations.}
\label{fig:Broccoli_vs_Tables}
\end{figure}

\xhdr{Multiple-choice results}
\Figref{fig:Broccoli_vs_Tables_MC} summarizes participants' vocabulary retention for all three conditions (pre-table, \B, post-table) on the MC test, both short-term (immediately after the learning section; dark colors) and long-term (1 to 4 weeks after the learning section; light colors).

We observe that learning with \B leads to significantly better short-term retention than table-based learning, with large gains in retention compared to both the pre-tables
($t(57)=6.5$, $p<10^{-7}$)
and post-tables
($t(57)=4.5$, $p<10^{-4}$).
This is particularly remarkable given that our study design deliberately disadvantages Broccoli by placing it between the two table conditions (as mentioned in \Secref{sec:Experimental design}, items shown first and last are remembered more easily than those shown in the middle \cite{murdock1962serial}).

In the long-term test, \B's performance is not significantly different from pre-tables
($t(31)=0.52$, $p=0.61$)
and post-tables
($t(31)=-0.19$, $p=0.85$),
at around 30\% correct answers, a 50\% increase over the random guessing baseline (20\%).
The drop in recall between the short- and long-term tests is expected, as the words were not revised between the two tests.
In a real deployment scenario, long-term retention would be increased via a spaced\hyp repetition\hyp based tutor algorithm (\cf\ \Secref{sec:Word scheduling}).%
\footnote{
Spaced repetition may be considered an orthogonal addition to \B as evaluated in this user study.
\Secref{sec:Compatibility} provides evidence that everyday information diets are indeed amenable to spaced repetition.
}

\begin{figure}[t]
\centering    
\subfloat[Multiple choice (short-term)]{\includegraphics[width=0.48\linewidth, keepaspectratio]{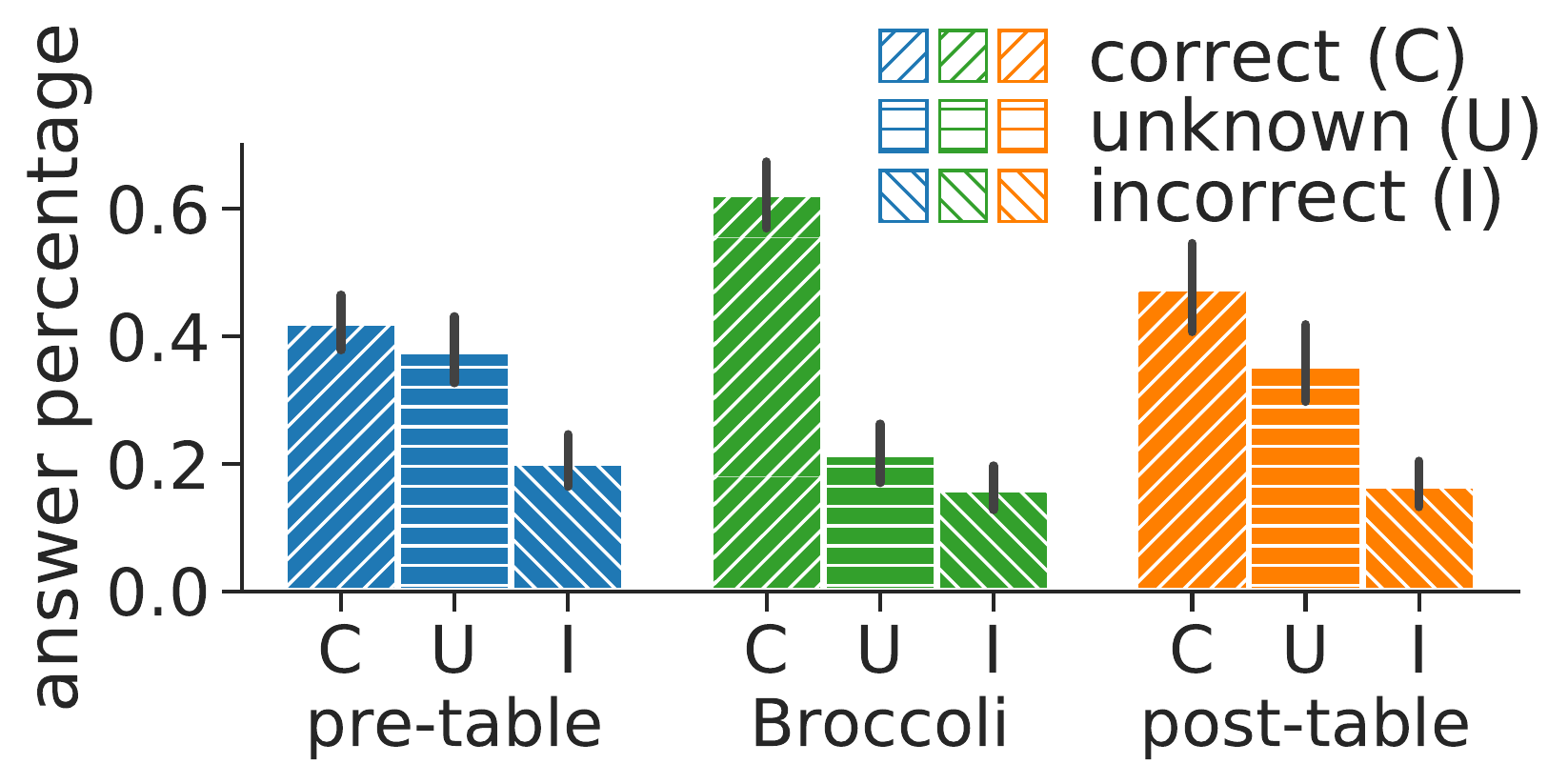}
}
\subfloat[Multiple choice (long-term)]{\includegraphics[width=0.48\linewidth, keepaspectratio]{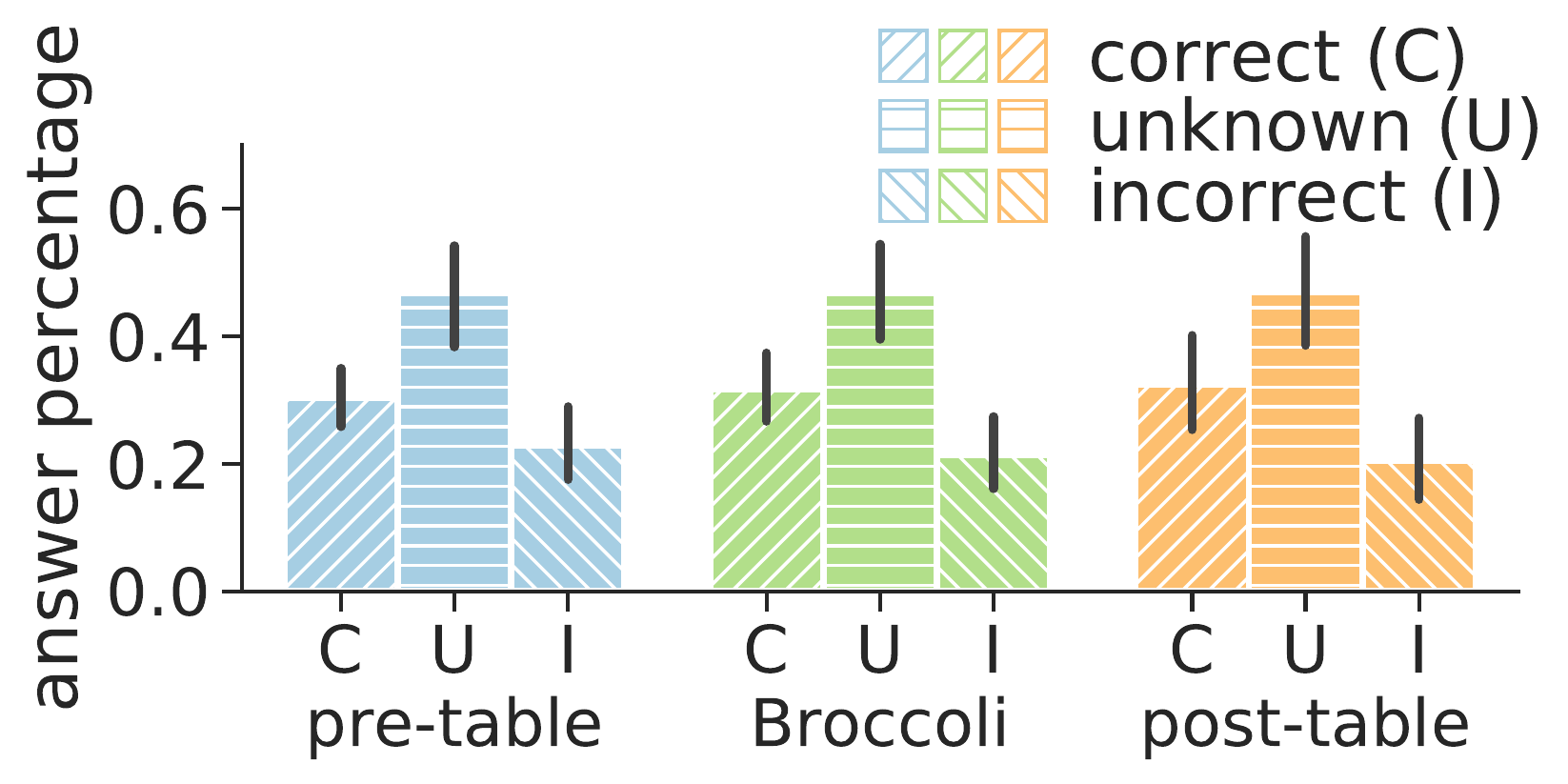}
}

\subfloat[Fill the gap (short\hyp term)]{\includegraphics[width=0.48\linewidth, keepaspectratio]{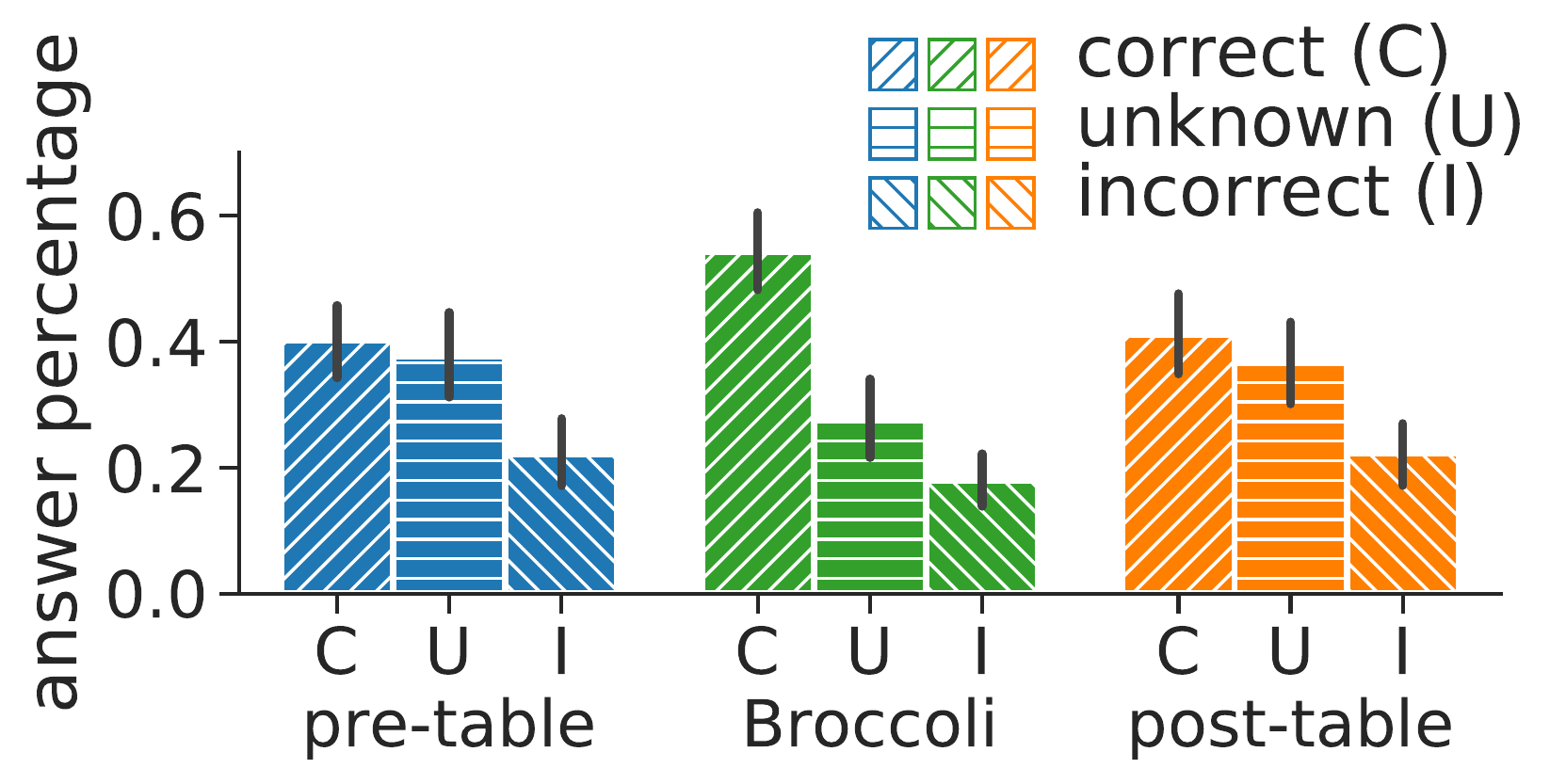}
}
\subfloat[Fill the gap (long-term)]{\includegraphics[width=0.48\linewidth, keepaspectratio]{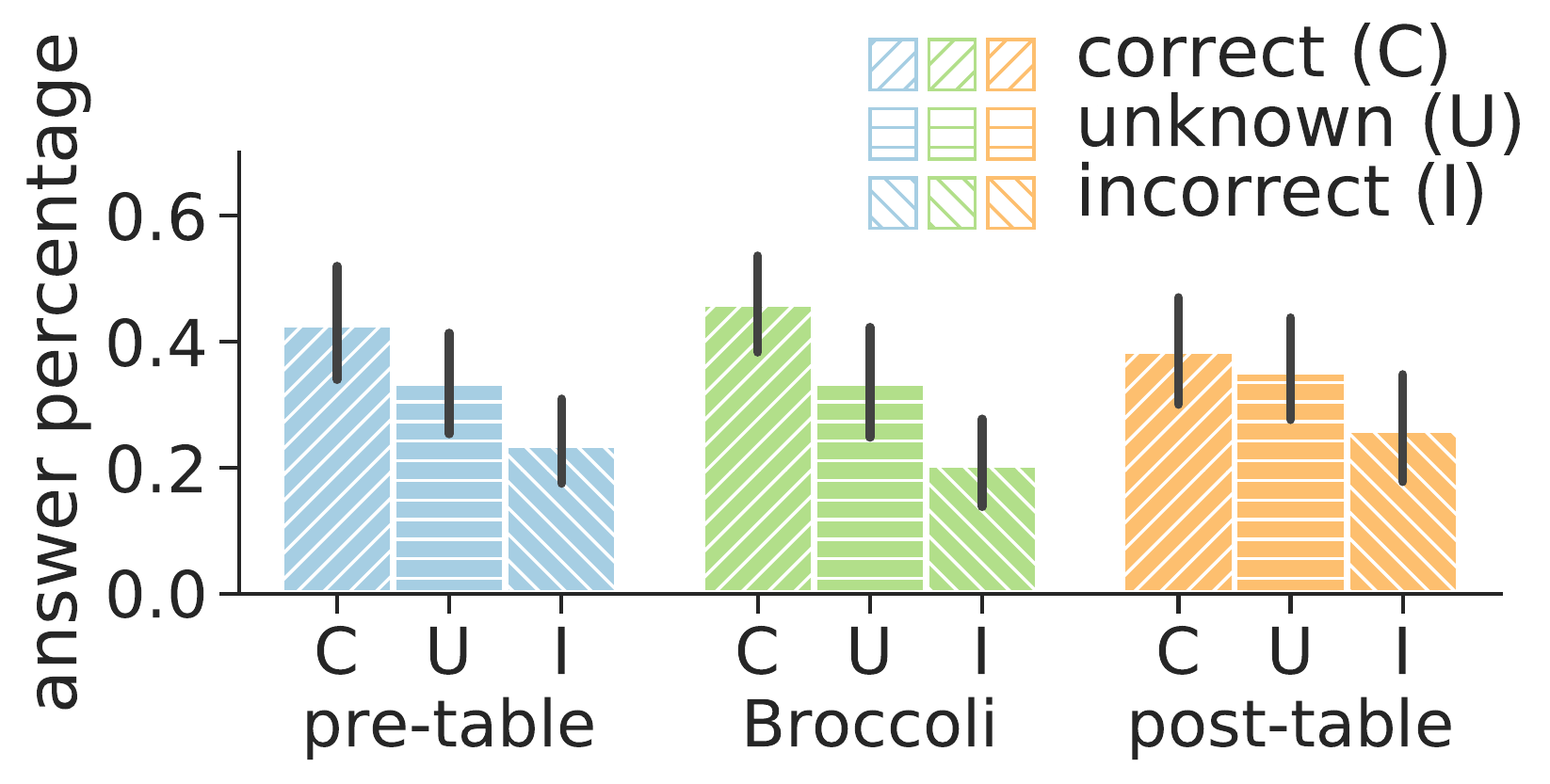}
}
\caption{
Broccoli \vs table\hyp based learning: Answer\hyp type percentages averaged over all users, with 95\% CIs.}
\label{fig:Broccoli_vs_Tables_detailed}
\end{figure}

Whereas \Figref{fig:Broccoli_vs_Tables_MC} computes the fraction of correct answers without distinguishing between ``incorrect'' und ``unknown'', for completeness we also show the distribution over all three answer types, including ``unknown'', in \Figref{fig:Broccoli_vs_Tables_detailed}a--b, which echoes the findings of \Figref{fig:Broccoli_vs_Tables_MC}.
In the remainder of the paper, we shall therefore mostly use the scalar performance metric (fraction of correct answers), as it is easier to aggregate and visualize.

\xhdr{Fill-the-gap results}
To evaluate the FTG test, we manually inspected all responses, labeling as correct those that were identical or synonymous to the respective English word before translation to Finnish. Differences in inflection (tense, singular \vs\ plural, \etc)\ and minor spelling mistakes were not counted as errors.
The results are shown in \Figref{fig:Broccoli_vs_Tables_FTG} and  \Figref{fig:Broccoli_vs_Tables_detailed}c--d.
As in the MC test, \B significantly outperforms pre-tables
($t(57)=4.4$, $p<10^{-4}$)
and post-tables
($t(57)=3.8$, $p<10^{-3}$)
in the short term, but not in the long term
(pre-tables: $t(28)=1.2$, $p=0.25$; post-tables: $t(28)=2.0$, $p=0.058$).
As the results are qualitatively equivalent to those from the MC test, we focus on the MC results from here on.

\begin{figure}[tb]
\centering    
\subfloat[Controlling for user]{\includegraphics[width=0.5\linewidth, keepaspectratio]{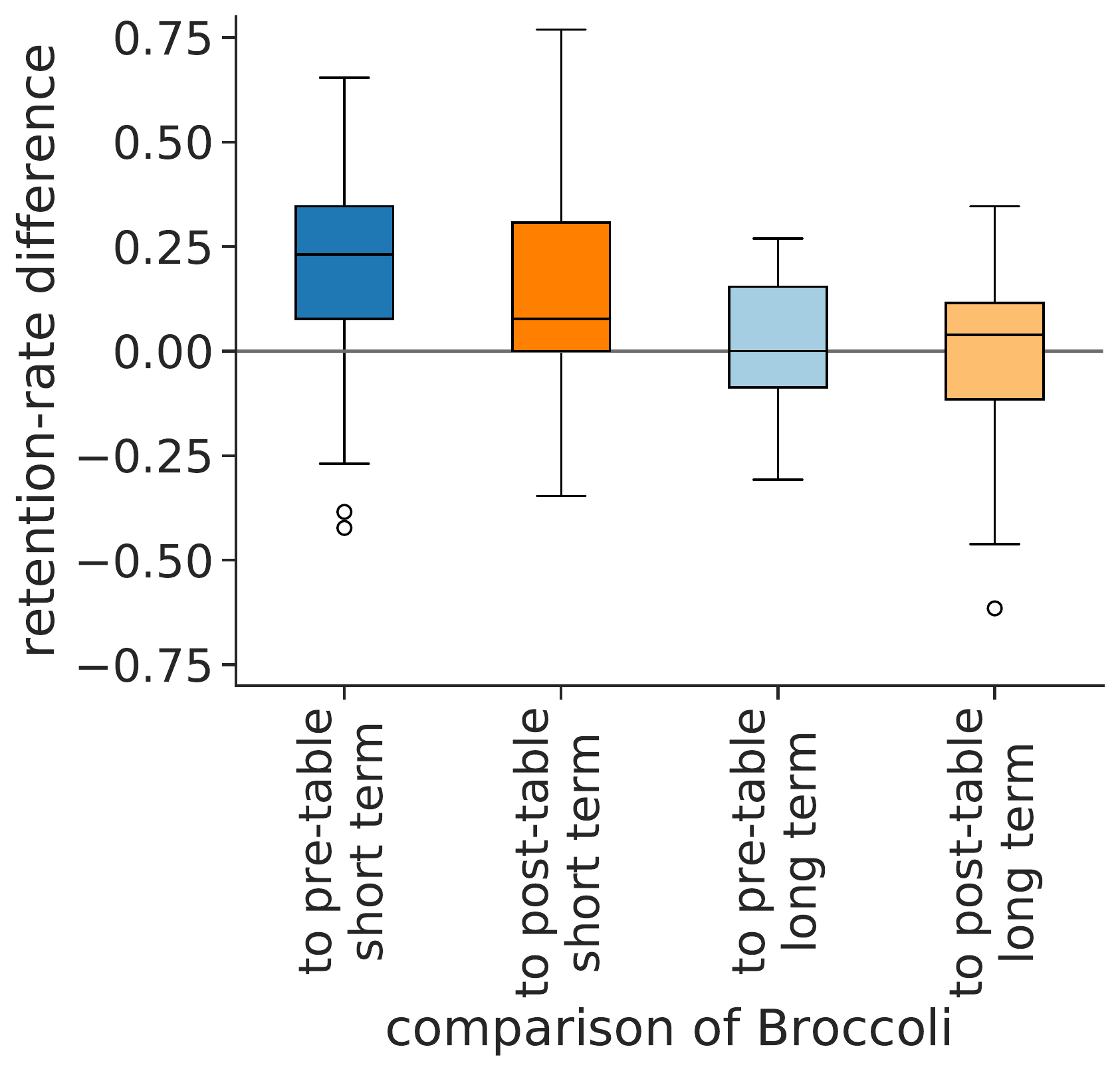}\label{fig:Broccoli_vs_Tables_controlling_for_user}}
\subfloat[Controlling for word]{\includegraphics[width=0.5\linewidth, keepaspectratio]{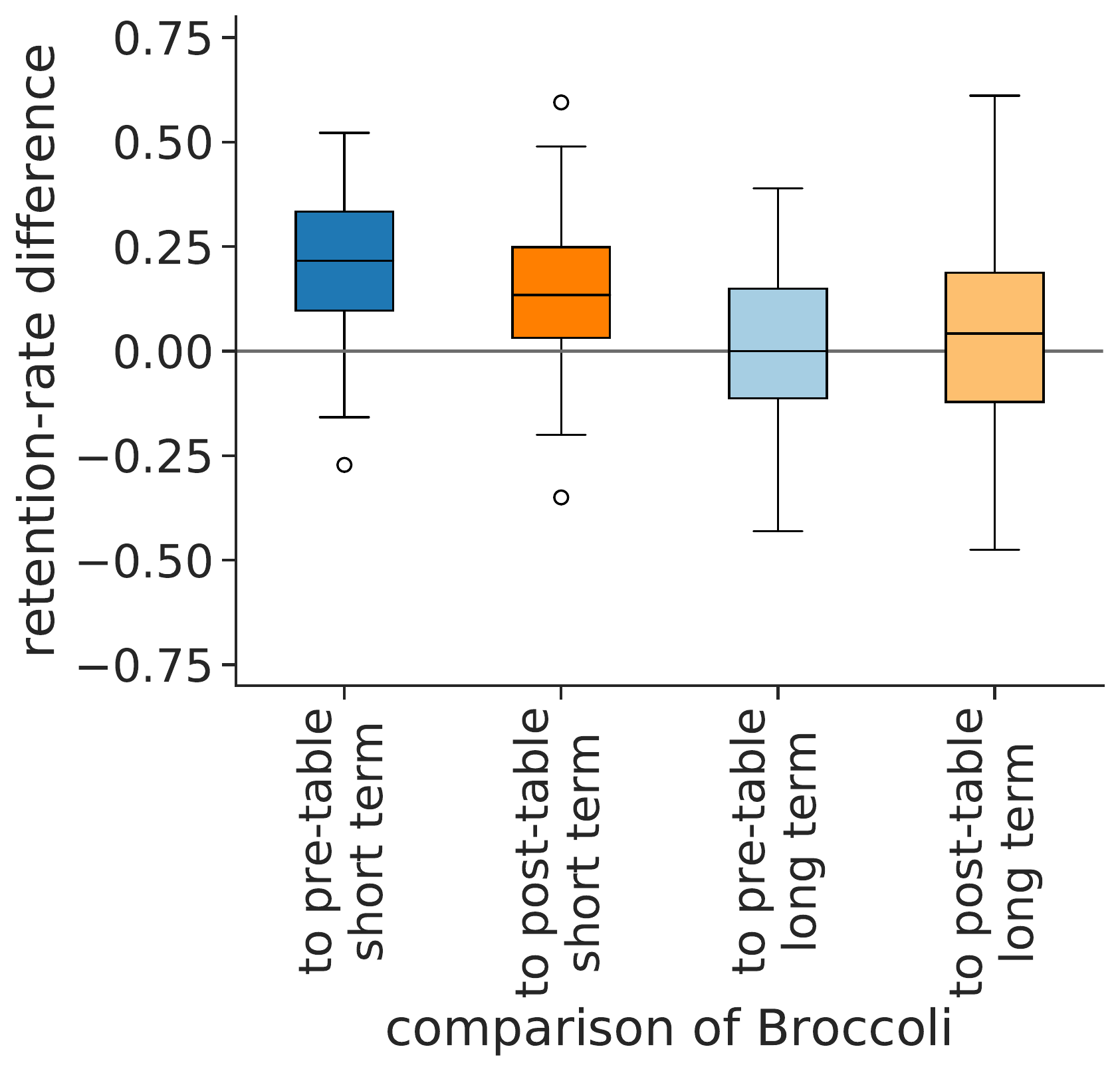}\label{fig:Broccoli_vs_Tables_controlling_for_word}}
\caption{
Broccoli \vs table-based learning:
Box plots of (a)~within-user and (b)~within-word differences in retention rates (\Secref{sec:Results RQ1}).
Positive values represent that \B performs better.
}
\label{fig:Broccoli_vs_Tables_controlling_for_user_and_word}
\end{figure}

\xhdr{Controlling for users and words}
We emphasize again that our randomized controlled within-subject design ensures that the results in all conditions (pre-table, \B, post-table) are comparable in a fair way, as each user and each word is evaluated in each condition.
This also allows for a more fine-grained analysis, where we evaluate the same user in each condition and compute the difference in performance between \B and the table conditions.
Similarly, we can fix a word and compute the difference in retention between \B and the table conditions.
The results indicate that \B leads to higher short-term retention for fixed users
(\Figref{fig:Broccoli_vs_Tables_controlling_for_user}; pre-tables: $t(57)=6.6$, $p<10^{-7}$; post-tables: $t(57)=4.5$, $p<10^{-4}$)
as well as for fixed words
(\Figref{fig:Broccoli_vs_Tables_controlling_for_word}; pre-tables: $t(77)=11$, $p<10^{-15}$; post-tables: $t(77)=7.4$, $p<10^{-9}$).
In the long term, the three conditions are again indistinguishable (all $p>0.46$).

\xhdr{Summary}
In a nutshell,
(1)~all evaluated learning conditions (\B and tables) outperform random guessing;
(2)~\B outperforms both table conditions short-term;
(3)~we cannot reject the null hypothesis that \B and the tables perform equally long-term.

\begingroup
\captionsetup[subfigure]{width=0.45\columnwidth}
\begin{figure*}[tb]
\centering    
\subfloat[Distribution of clicks per word]{\includegraphics[width=0.5\columnwidth, keepaspectratio]{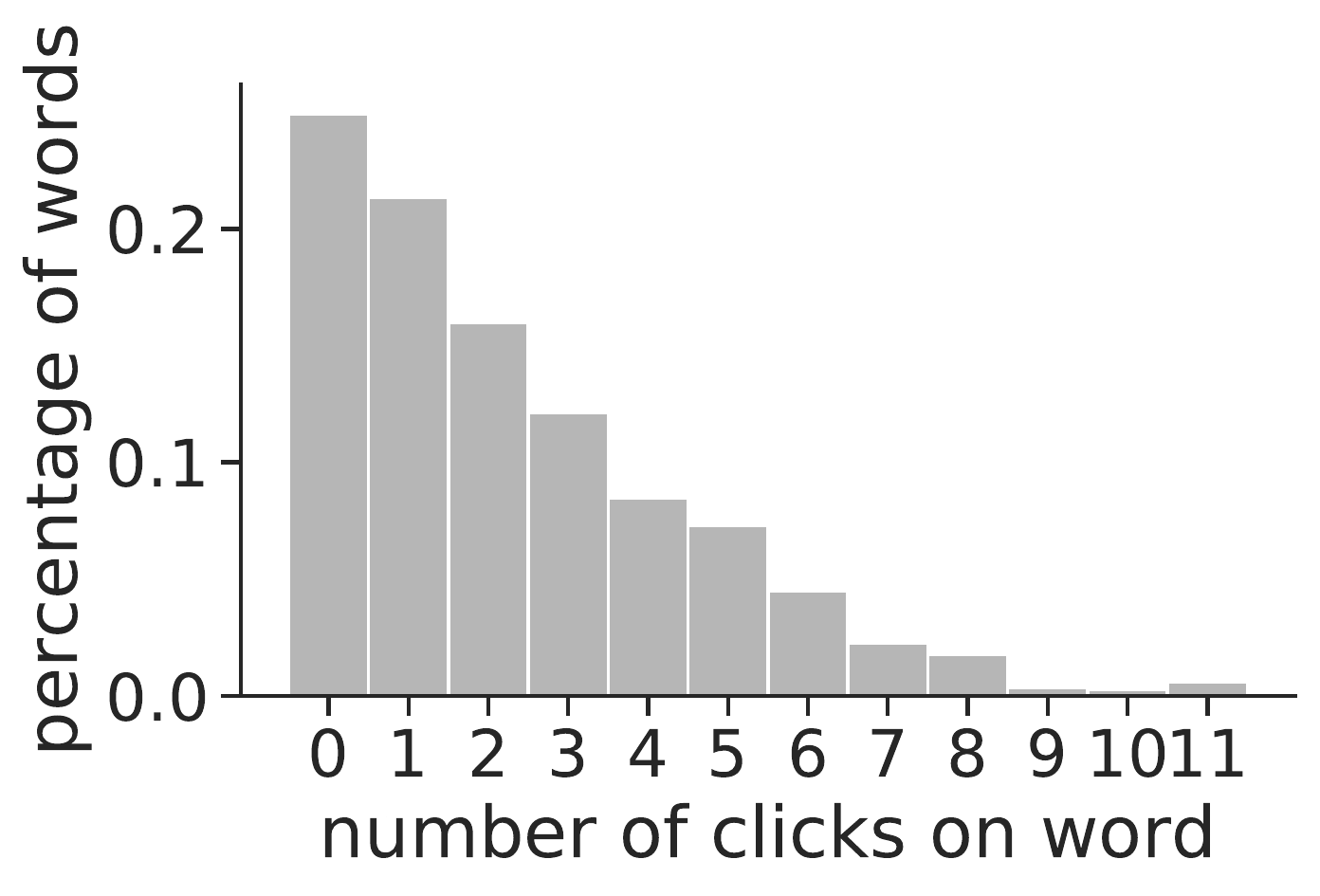}\label{fig:histogram_number_of_clicks}}
\subfloat[%
Word easiness \vs clicks%
]{\includegraphics[width=0.5\columnwidth, keepaspectratio]{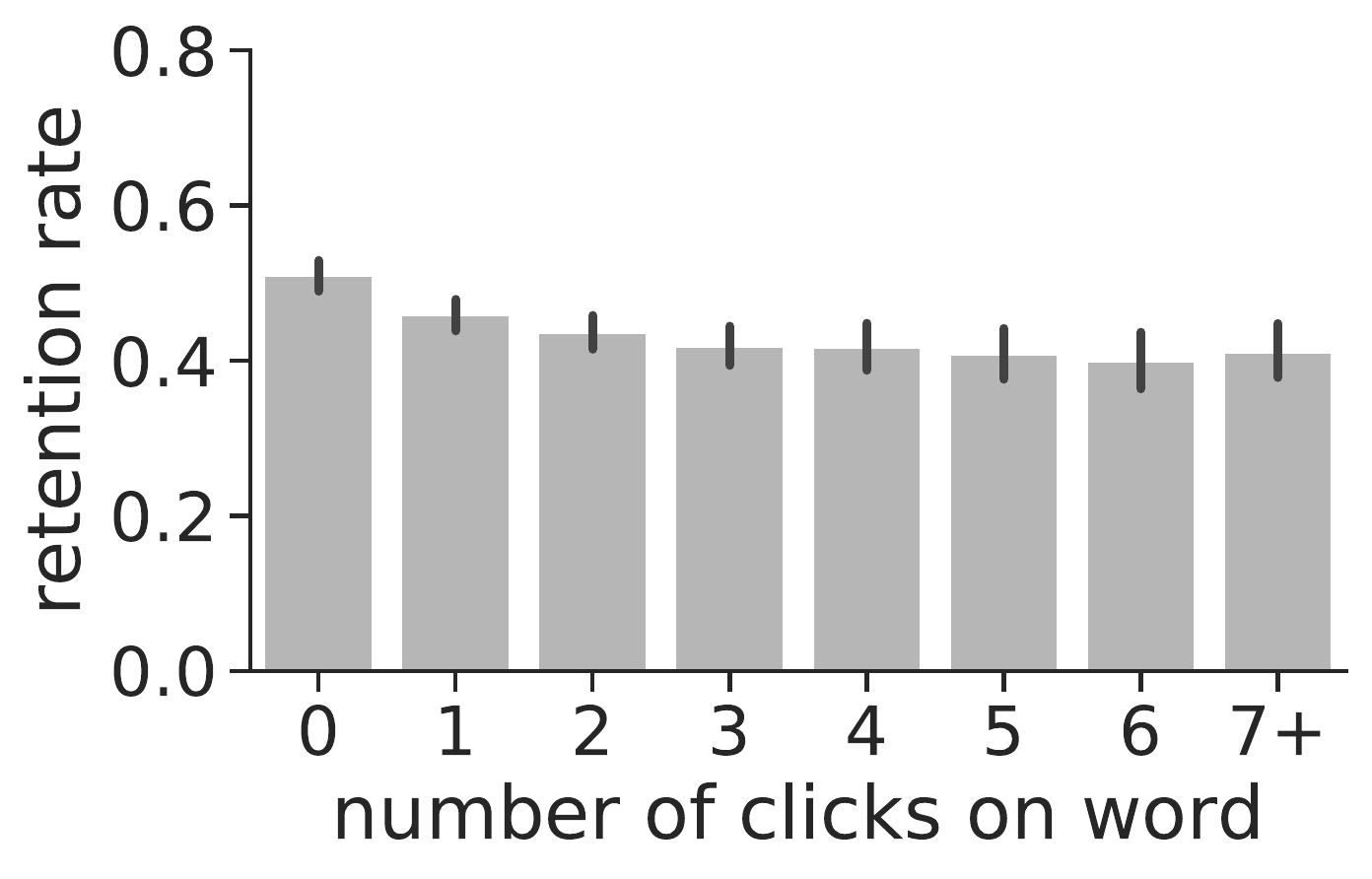}\label{fig:clicks_word_difficulty}}
\subfloat[%
Results \vs clicks (short-term)%
]{\includegraphics[width=0.5\columnwidth, keepaspectratio]{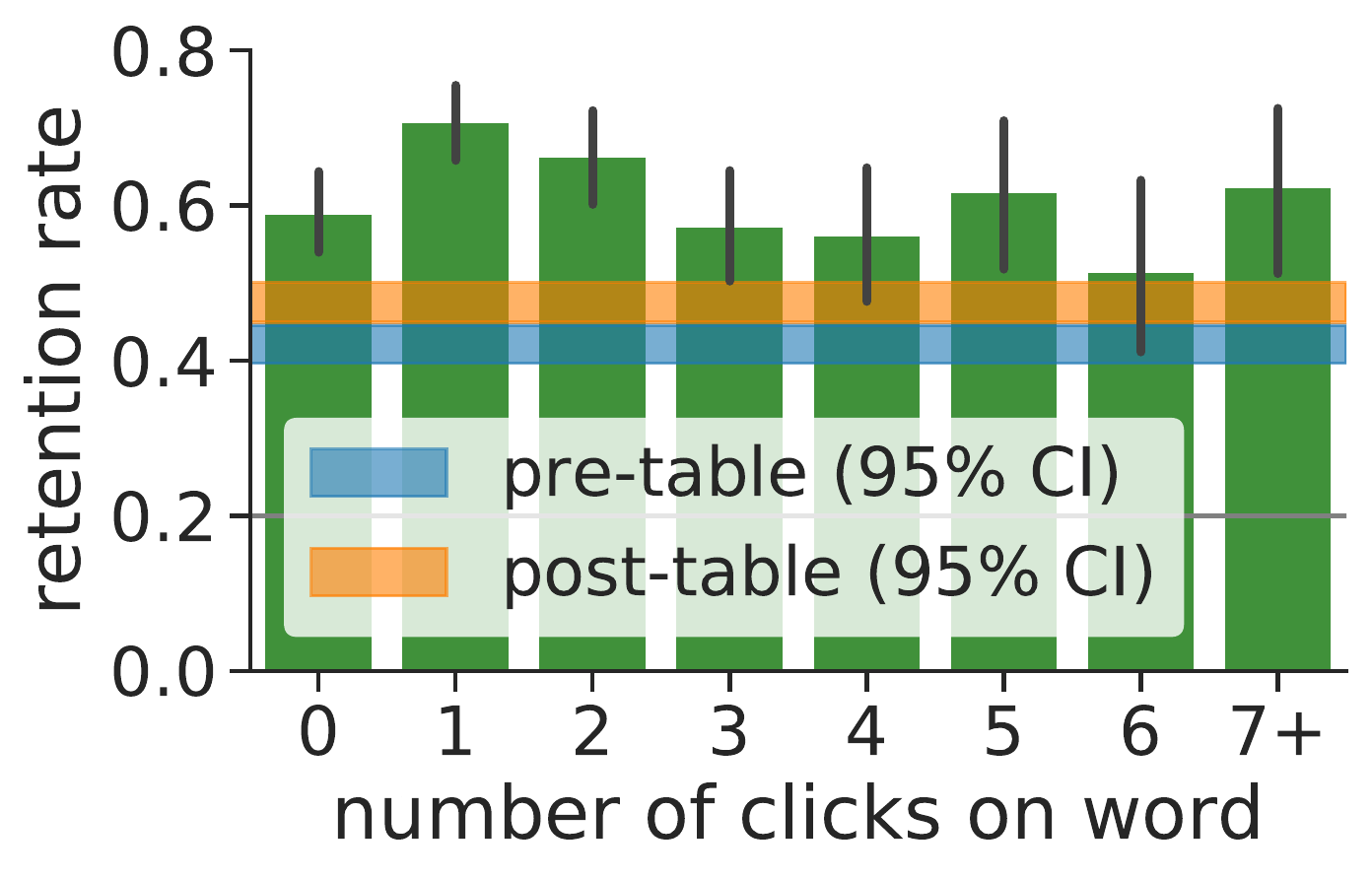}\label{fig:impact_clicking_short_term}}
\subfloat[%
Results \vs clicks (long-term)%
]{\includegraphics[width=0.5\columnwidth, keepaspectratio]{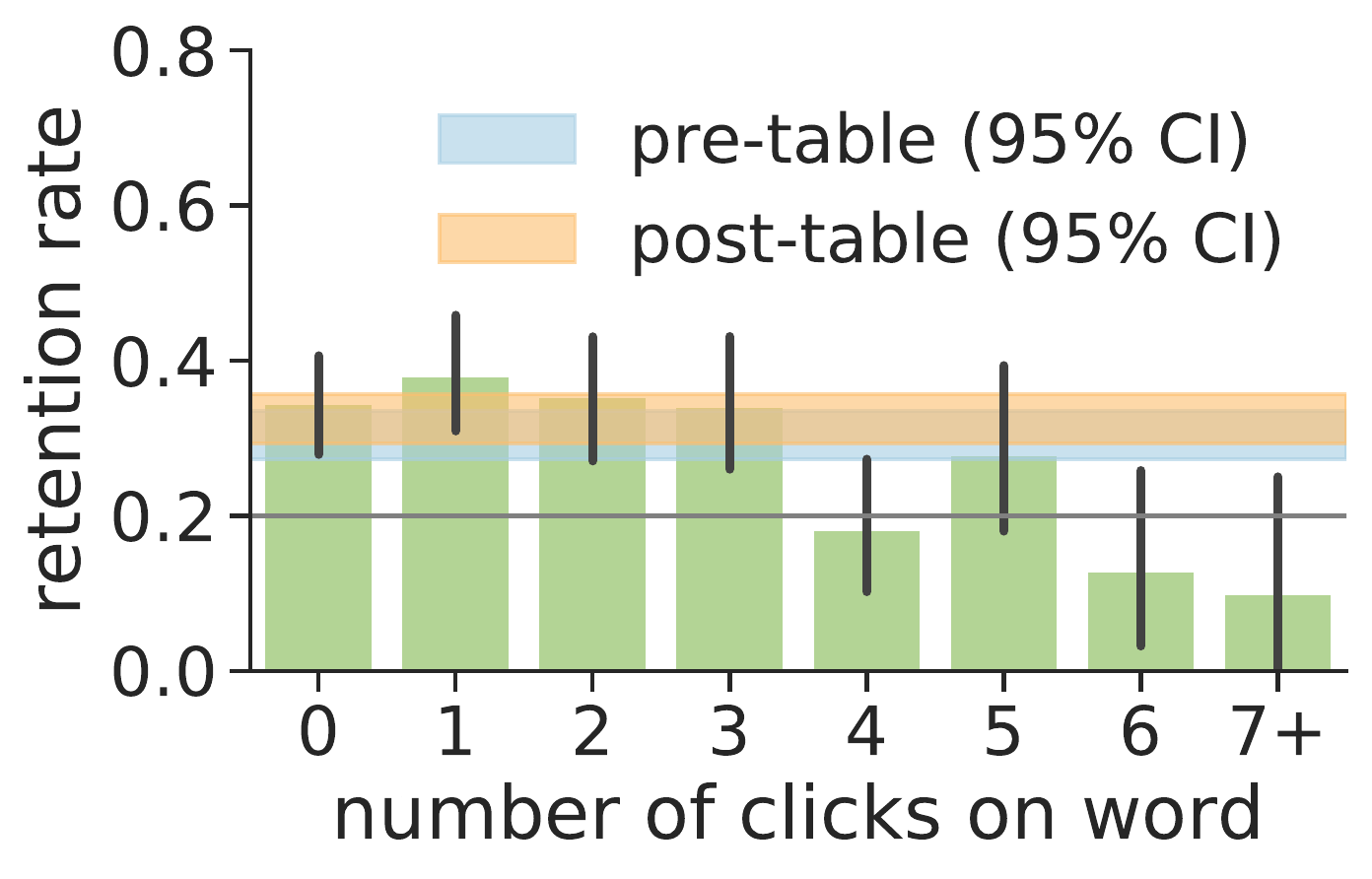}\label{fig:impact_clicking_long_term}}
\caption{
Impact of revealing translations by clicking on words in \B condition:
(a)~Distribution of clicks per word, averaged over all users.
(b)~Word easiness (measured as word's retention rate in table conditions).
(c)~Retention rates in short-term multiple\hyp choice test.
(d)~Retention rates in long-term multiple\hyp choice test.
Errors bars are 95\% CIs; gray horizontal lines in (c) and (d): random guessing among 5 candidate translations.
}
\label{fig:clicking}
\end{figure*}
\endgroup

\subsection{RQ2: Impact of revealing translations}
\label{sec:Results RQ2}

Recall that the \B prototype tested in the user study allowed users to reveal the original word (before it was translated into the target language) by clicking on the translated word.
We now investigate the impact of such clicks on word retention.

As explained, in the \B condition each word appeared in 7 instances for each user, and each instance could be clicked zero or more times.
\Figref{fig:histogram_number_of_clicks} plots the distribution of clicks per word--user pair (the 7 instances of the same word are counted toward the same word--user pair).
The probability of clicks per word is strictly decreasing, and in particular, one quarter of the words are never clicked by the average user, an indicator that even in a lab setting where users are aware that they will be tested on word retention, \B's ideal usage mode of learning without explicit interaction comes naturally to users. We expect that in a real-world usage scenario, the fraction of words whose translation is never requested would be even higher.

\Figref{fig:clicks_word_difficulty} plots the easiness of words as a function of the number of clicks. Here we measure the easiness of a word as the fraction of users who learned the word in one of the table conditions and recalled it correctly in the short-term test (but clicks are counted in the \B condition, of course).
We hypothesized that easy-to-guess words would be clicked less frequently, and indeed this is confirmed by \Figref{fig:clicks_word_difficulty}, but the effect is subtle, with easiness decreasing only by 10 percentage points over the full range of clicks.

Ideally, \B users would not need to see any ``solutions'' of correct translations in order to successfully retain vocabulary.
We investigate if this ideal is feasible in practice by plotting the retention rate for words learned in the \B condition as a function of the number of clicks the respective user made on them, for both the short-term (\Figref{fig:impact_clicking_short_term}) and long-term (\Figref{fig:impact_clicking_long_term}) test.
We see that the retention rate does not systematically vary with the number of clicks and that, in the short-term test (\Figref{fig:impact_clicking_short_term}), even words that were never clicked are retained by the respective users at a rate significantly higher than words learned in either table condition.
In the long-term test (\Figref{fig:impact_clicking_long_term}), too, words that were seen in the \B condition and clicked between 0 and 4 times are indistinguishable with respect to retention, and they are recalled at about the same rate as words learned in the table conditions.
These are among the most important results of our study, as they establish that \B can teach vocabulary at competitive rates purely via passive exposure in context, without any explicit user interaction.

\subsection{RQ3: \B-induced slowdown}
\label{sec:Results RQ3}

\begin{figure}[t]
    \resizebox{\linewidth}{!}{
        \includegraphics{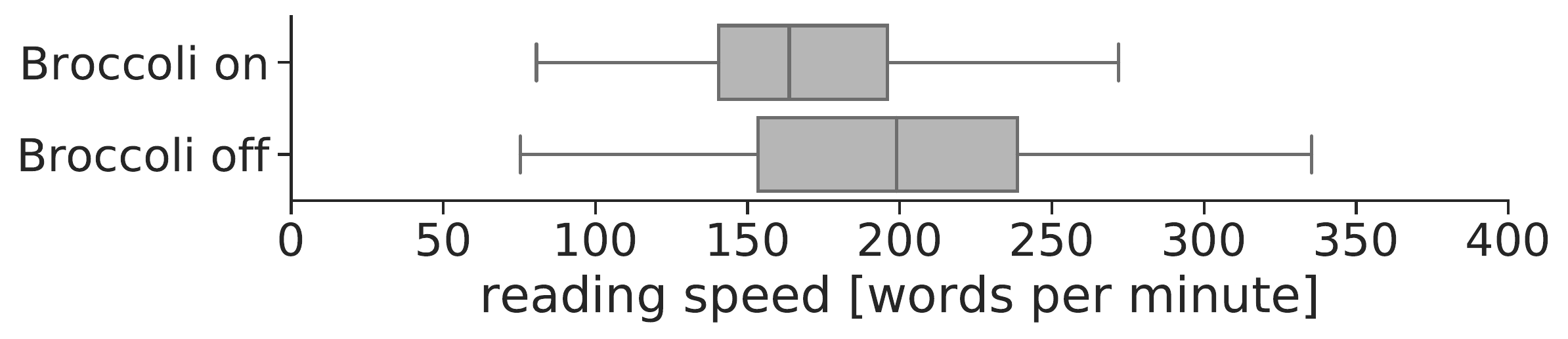}
    }  
    \caption{
    Broccoli-induced slowdown:
    Reading speed with \vs without \B.}
    \label{fig:reading_speed_comparison}
\end{figure}

We expect the fact that text in the \B condition contains words in a foreign language to decrease users' reading speed.
As described in \Secref{sec:Experimental design}, we designed the study to allow for quantifying this effect: Next to the article snippets augmented with translated words, the \B section also included one page into which no translations were inserted. On this unprocessed page, we can estimate the user's natural reading speed.
\Figref{fig:reading_speed_comparison} summarizes the distribution over all users' reading speeds with and without \B.
Users' median reading speed is 199 words per minute without \B, \vs\ 163 with \B, implying a \B-induced slowdown of 18\%.
Note that this does not necessarily imply a \emph{conscious} slowdown, such as would be indicated by the usage of mnemonic strategies for memorizing words. Indeed, in \Secref{sec:mnemonics} we will see that such strategies are applied extremely rarely in the \B condition (we will also discuss the slowdown in \Secref{sec:Discussion}).

\subsection{RQ4: Impact of language model}
\label{sec:Results RQ4}

\begin{figure}[tb]
\centering
\begin{minipage}{0.48\columnwidth}
    \centering
    \includegraphics[width=\textwidth, keepaspectratio]{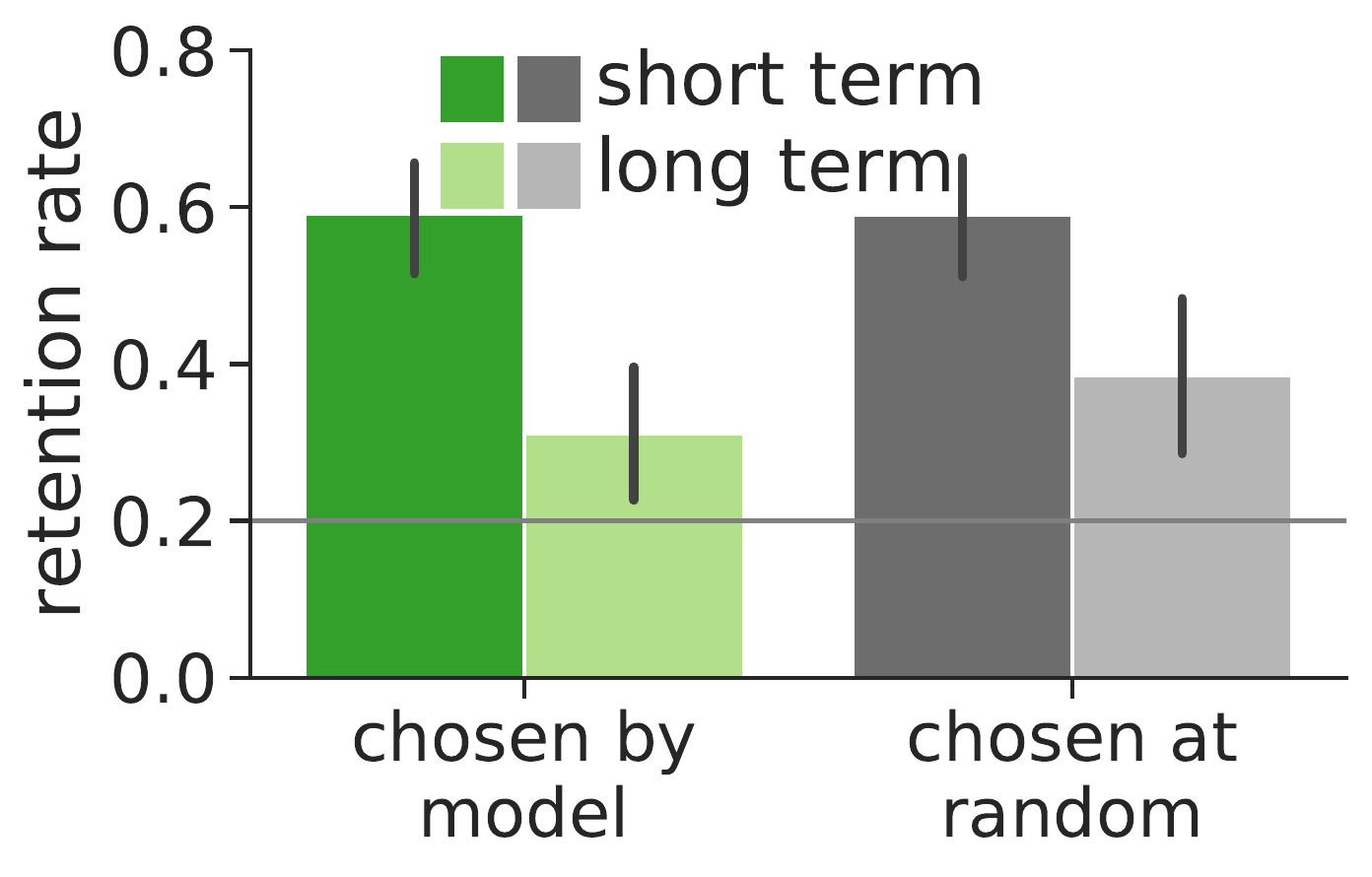}
    \caption{Impact of language model on retention rate.}\label{fig:impact_language_model}
\end{minipage}\hfill
\begin{minipage}{0.48\columnwidth}
    \centering
    \includegraphics[width=\textwidth, keepaspectratio]{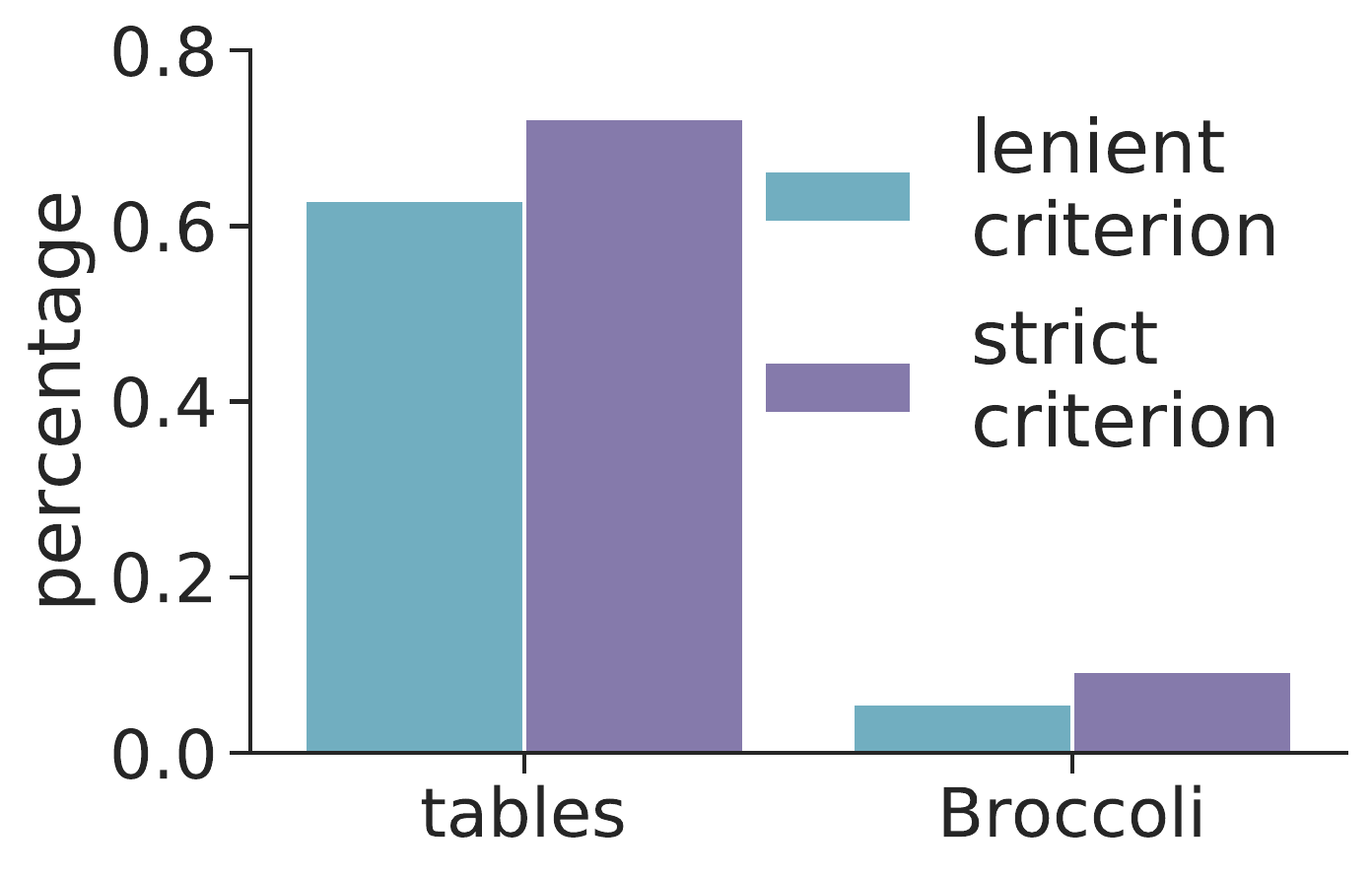}
    \caption{Mnemonic\hyp strategy use in tables \vs Broccoli.}\label{fig:strategy_use}
\end{minipage}
\end{figure}

Next, we investigate the effect of using a language model when selecting words for translation (\cf \Secref{sec:Context-based word scoring}).
The goal of the language model is to identify instances where the given word appears in a typical context, which we expected to lead to increased word retention.
As seen in \Figref{fig:impact_language_model}, this is not the case: words translated in contexts that scored highly according to the language model (green, left) have a retention rate indistinguishable from that of words translated in random contexts (gray, right).
We further discuss potential causes of this effect, as well as the (positive) implications of the language model's weak role, in \Secref{sec:Discussion}.

\subsection{RQ5: Use of mnemonic strategies}
\label{sec:Results RQ5}
\label{sec:mnemonics}

The ideal followed by \B is to avoid explicit cognitive efforts on the learner's behalf and enable vocabulary learning in a purely passive fashion.
A common strategy in classical, memorization\hyp based vocabulary learning (\eg, via flashcards) is to invent ad-hoc mnemonic devices, or ``memory hooks'', that facilitate memorization by establishing salient phonological, visual, semantic, \etc, connections between the foreign word to be memorized and words in the learner's mother tongue.%
\footnote{
One participant in our study reported this mnemonic device for Finnish \textit{karhu} (\textit{bear}): ``\textit{kar} sounds like \textit{car,} \textit{hu} is close to \textit{hood.} I imagine a bear ripping off a car hood.''
}
While effective \cite{bellezza1981mnemonic}, such strategies require self\hyp discipline and creativity, and come at a high cognitive cost. If foreign\hyp language vocabulary learning could be achieved without them, it would become a more attractive activity for anyone---which is the very philosophy behind \B.

We therefore conclude our study evaluation by analyzing the prevalence of mnemonic devices in the \B and table conditions.
The exit survey of the study included the two questions, ``Did you use any tricks or strategies to remember translations (1)~from the vocabulary tables? (2)~when reading the Wikipedia articles?'', asking for a free-text response where participants could describe the tricks or strategies they had used in each condition.
We manually coded the responses into two classes, according to whether they had or had not employed mnemonic strategies.

We expected a lower prevalence of mnemonic strategies in \B, compared to the table conditions.
\Figref{fig:strategy_use} shows how strikingly this expectation bears out in practice:
the prevalence of mnemonic strategies in table-based learning is at least 7 times as high as in \B (namely, when applying a strict criterion for counting mnemonic strategies to \B, and a lenient criterion to tables), at 63--72\% for tables and 6--9\% for \B.

These results, together with \B's competitive retention rates (\Secref{sec:Results RQ1}), present a strong argument for the paradigm advocated in this paper:
in-context vocabulary learning is possible with very little unnatural memorization effort and no loss in retention.


\section{Compatibility of Broccoli with conventional information diets}
\label{sec:Compatibility}

In the above user study, the texts shown to users were compiled into a predetermined sequence specifically such that each translated word was seen by the user multiple times (7 times).
In a real setting, the sequence of texts would be determined not by us, but by the user's natural information diet, and it would be the role of the tutor algorithm (\Secref{sec:Word scheduling}) to make sure that words are revised (translated) in a rhythm that allows them to be effectively retained.

\B can be expected to work well if the words that the tutor algorithm wants to translate at a specific time occur in the text that the user reads at that time.
If all words were frequent (which is of course not the case) any word could be scheduled for repetition at any time.
Hence, when attempting to assess how realistically \B can be applied to natural information diets outside of a lab setting, the crux is to determine how badly the tutoring algorithm's natural rhythm is hampered by infrequently occurring words.

In the following, we first introduce our methodology for answering this question, assuming we are given a corpus that the user reads top to bottom.
Then we explain how we generate corpora and discuss the results obtained on them (\Secref{sec:Wikipedia reading} and \ref{sec:Book reading}).

\xhdr{Lemma revisitation time}
In order to quantify how much the tutor algorithm is constrained by infrequent lemmas,%
\footnote{We operate on lemmas rather than raw tokens because we do not want to distinguish different inflected forms of the same word. Punctuation and numbers are discarded before the analysis.}
we proceed as follows.
Given a corpus and a lemma $x$,
(1)~count the number of tokens between all subsequent occurrences of $x$,
(2)~translate each token distance into a time difference by assuming a reading speed of 200 words per minute (\cf\ \Figref{fig:reading_speed_comparison}) and a reading time of 3 hours per day, and
(3)~define the \textbf{revisitation time} of $x$ as the 90th percentile of the time differences for $x$.
Finally, we aggregate lemma\hyp specific revisitation times into a corpus\hyp wide revisitation time via the 90th percentile over the set $\mathcal{L}$ of all lemmas that are to be taught.
Intuitively, a corpus\hyp wide revisitation time of $t$ days implies that nearly all subsequent occurrences of nearly all lemmas appear within $t$ days of one another.
Revisitation times depend linearly on reading speed and daily reading time: for a user who reads half as fast as our assumed reading speed, or spends half as much time reading, revisitation times would double.

In our analysis, we work with various lemma sets $\mathcal{L}_\alpha$, where $\mathcal{L}_\alpha$ is defined as the smallest set of most frequent lemmas accounting for a fraction $\alpha \in [0,1]$ of all tokens in the corpus.
(We hence call $\alpha$ the \textbf{corpus coverage.})
The intuition is that after learning the translations of the lemmas $\mathcal{L}_\alpha$ in the target language, a user would understand a fraction $\alpha$ of a similar corpus in the target language.

\xhdr{Scenarios}
In the following, we consider two common scenarios,
Web browsing (\Secref{sec:Wikipedia reading}) and
fiction e-book reading (\Secref{sec:Book reading}).
We consider English as a source language, but given the universality of statistical laws of language \cite{zipf1949human}, we anticipate the results to be similar for different source languages.

\begin{figure}[tb]
\centering    
\subfloat[Lemma revisitation time]{\includegraphics[width=0.5\columnwidth, keepaspectratio]{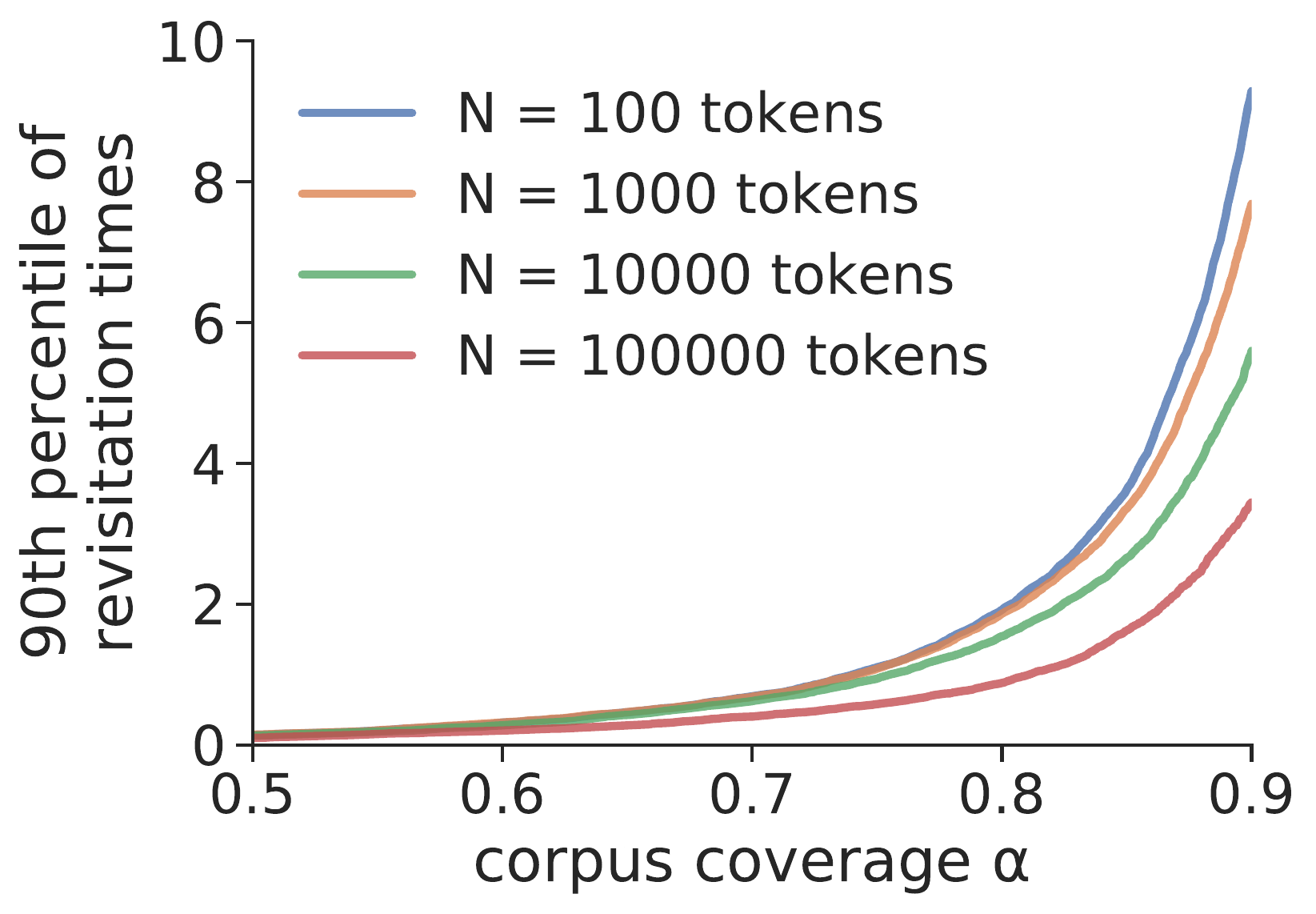}\label{fig:random_walk_p90p90}}
\subfloat[Vocabulary size]{\includegraphics[width=0.5\columnwidth, keepaspectratio]{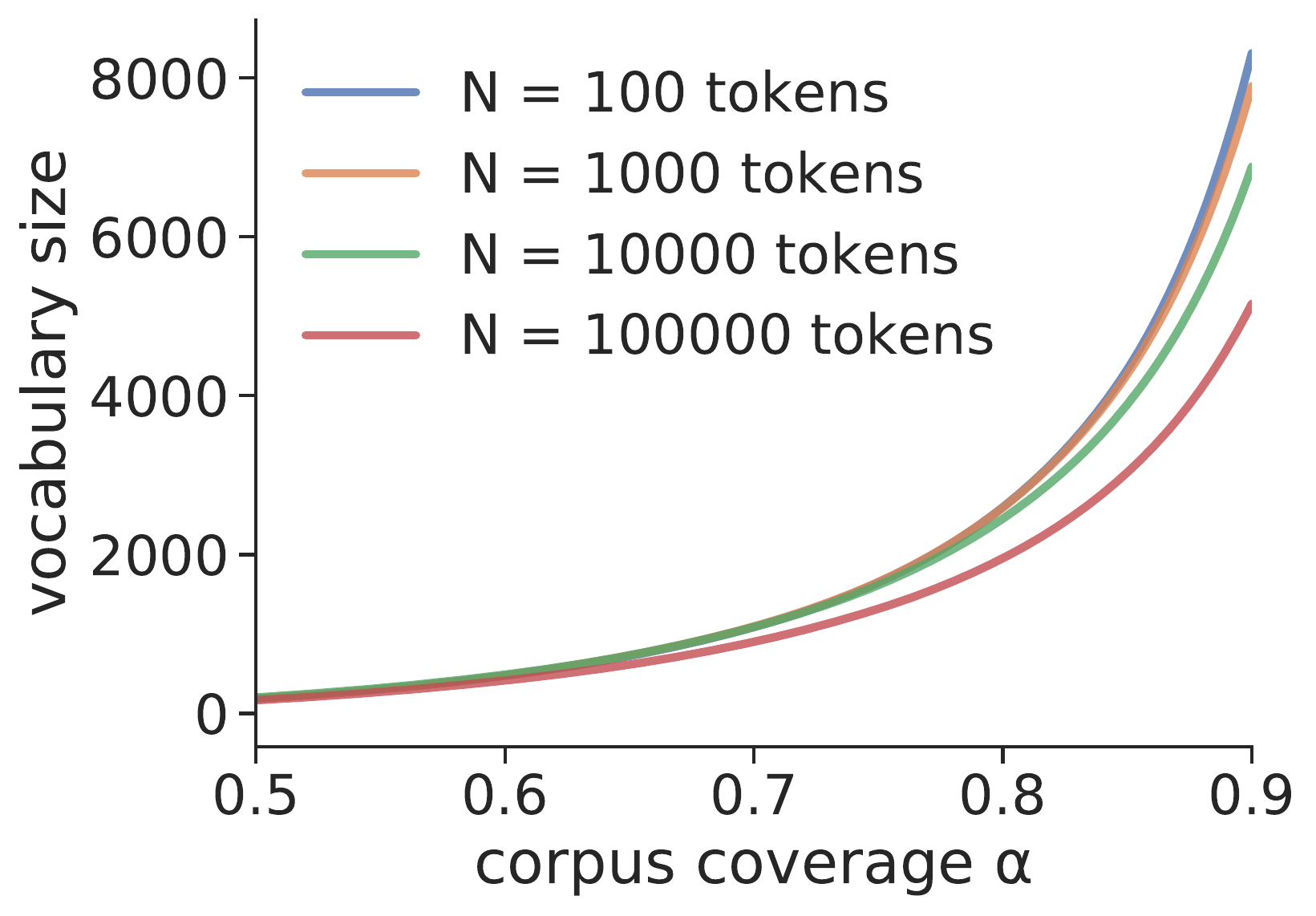}\label{fig:random_walk_vocabsize}}
\caption{Compatibility of \B with Wikipedia browsing:
Growth of
(a)~lemma revisitation time and
(b)~vocabulary size
as functions of corpus coverage $\alpha$.
}
\label{fig:wiki_walks}
\end{figure}

\subsection{Web browsing}
\label{sec:Wikipedia reading}

In the first scenario, the corpus consists of the concatenation of Web pages consumed by a user in temporal order.
As real Web browsing histories are hard to obtain, and considering that Wikipedia is one of the world's most prominent websites, covers a wide variety of topics, and comes with publicly available statistics about its usage, we focus on the special case of (simulated) Wikipedia browsing.

\xhdr{Corpus generation}
In order to generate Wikipedia browsing histories, we use the publicly available Clickstream data \cite{Wulczyn2017}, which specifies the click frequency of every Wikipedia link $(u,v)$, as well as the frequency of visits to $u$ without an ensuing click.
The Clickstream data allow us to generate synthetic yet realistic \textbf{browsing sessions} via biased random walks according to empirically estimated transition probabilities.%
\footnote{It has been shown that first-order Markov chains, as defined by the Clickstream, are good models of Web browsing, including Wikipedia \cite{singer2014detecting}.}
The starting page $u_0$ of a session is sampled with probability proportional to $u_0$'s overall visit count.
From there, we perform a biased random walk that, at any step, stops with the present page's no-click probability, in which case we jump back to $u_0$ and start a fresh walk from there.
The goal of repeatedly restarting from $u_0$ is to generate topically coherent reading sessions containing an arbitrarily large number $N$ of tokens.
Once the concatenation of pages in a session surpasses $N$ tokens, we terminate the session%
\footnote{When concatenating walks into sessions, we include $u_0$ only the first time.}
and repeat the process (from a newly sampled $u_0$) until the concatenation of all sessions reaches 2 million tokens.
The resulting sequence of 2 million tokens then serves as the corpus.

By varying $N$, we can model users with varying breadth of interests (smaller $N$ implies more frequent topic switching), which is useful as topical breadth might affect the applicability of \B.

\xhdr{Results}
The lemma revisitation time for simulated Wikipedia reading is shown in \Figref{fig:random_walk_p90p90} for various combinations of corpus coverage $\alpha$ and session length $N$.
As rarer words are included in the vocabulary to be learned (larger $\alpha$), and as the reading history becomes more topically diverse (smaller $N$), revisitation times increase.
Remarkably though, even in the most extreme setting ($\alpha=0.9$, $N=100$), nearly all lemmas (namely, 90\% of them) are typically (namely, 90\% of the time) encountered at least every 10 days.
In less extreme settings (smaller $\alpha$, larger $N$), revisitation times drop below 2 days.
This regularity of occurrences implies that the vast majority of lemmas can be revised shortly after the tutor algorithm schedules them for revision.
Additionally, the large vocabulary sizes (thousands of lemmas, \Figref{fig:random_walk_vocabsize}) imply that any text will presumably contain many lemmas that the tutoring algorithm currently considers worth revising.
In practice, the tutor algorithm will be faced with an overabundance of word occurrences. In the extreme case of $\alpha=0.9$ and $N=100$, using the revisitation time of 10 days to cycle just once through the 8,000 unique words would require the user to review an average of 800 words each day. 
Together, these findings confirm the practical applicability of \B for Wikipedia information diets.

\subsection{Fiction e-book reading}
\label{sec:Book reading}

Fiction books constitute a relevant application domain for \B because here texts are topically coherent, which presumably facilitates learning in context.
They are also well suited for evaluating the real-world feasibility of \B, since the exact sequence in which words will be encountered is predetermined, such that real reading histories are readily available and need not be simulated.

\begin{figure}[tb]
\centering    
\subfloat[Lemma revisitation time]{\includegraphics[width=0.5\columnwidth, keepaspectratio]{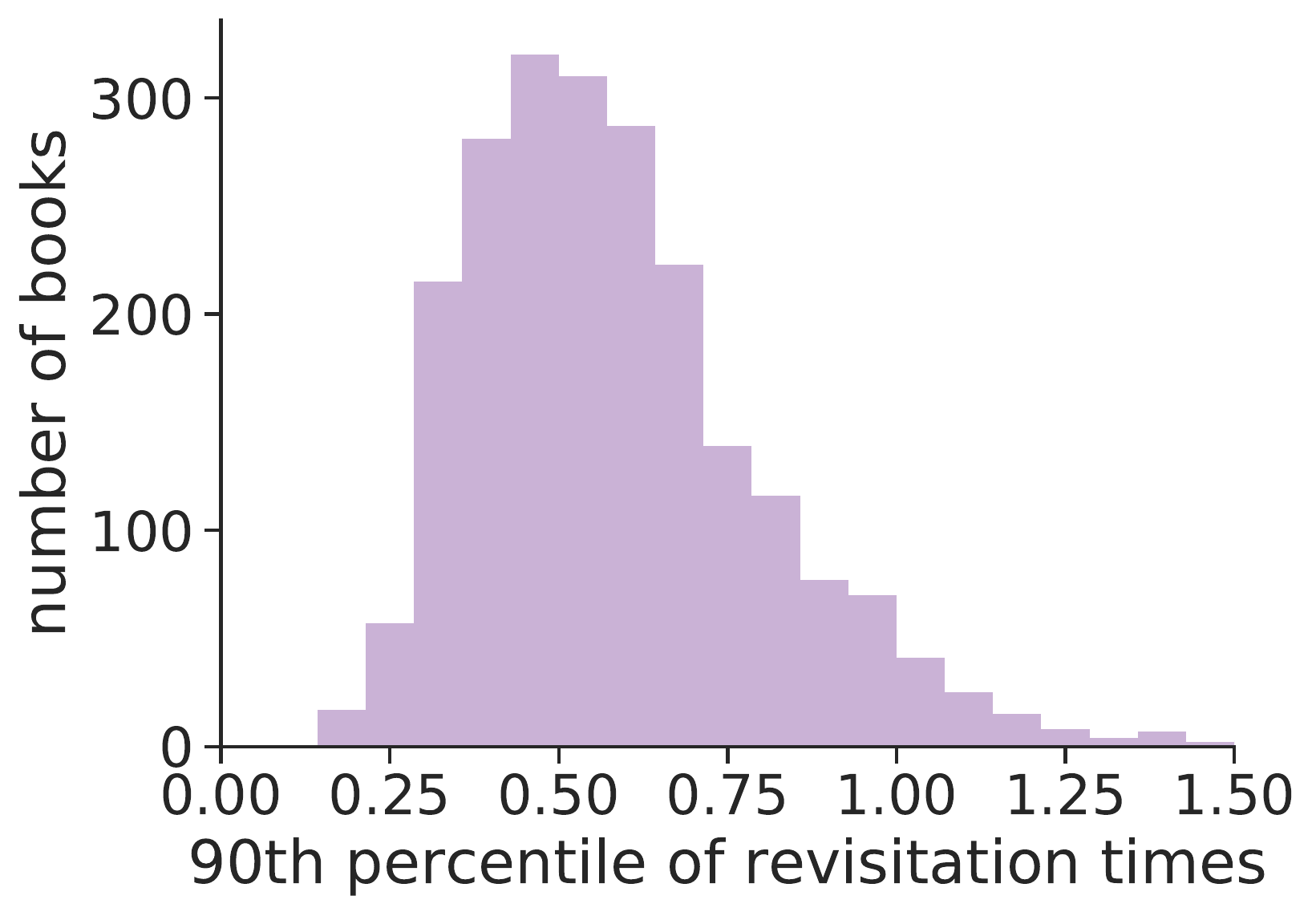}\label{fig:gutenberg_distplot}}
\subfloat[Vocabulary size]{\includegraphics[width=0.5\columnwidth, keepaspectratio]{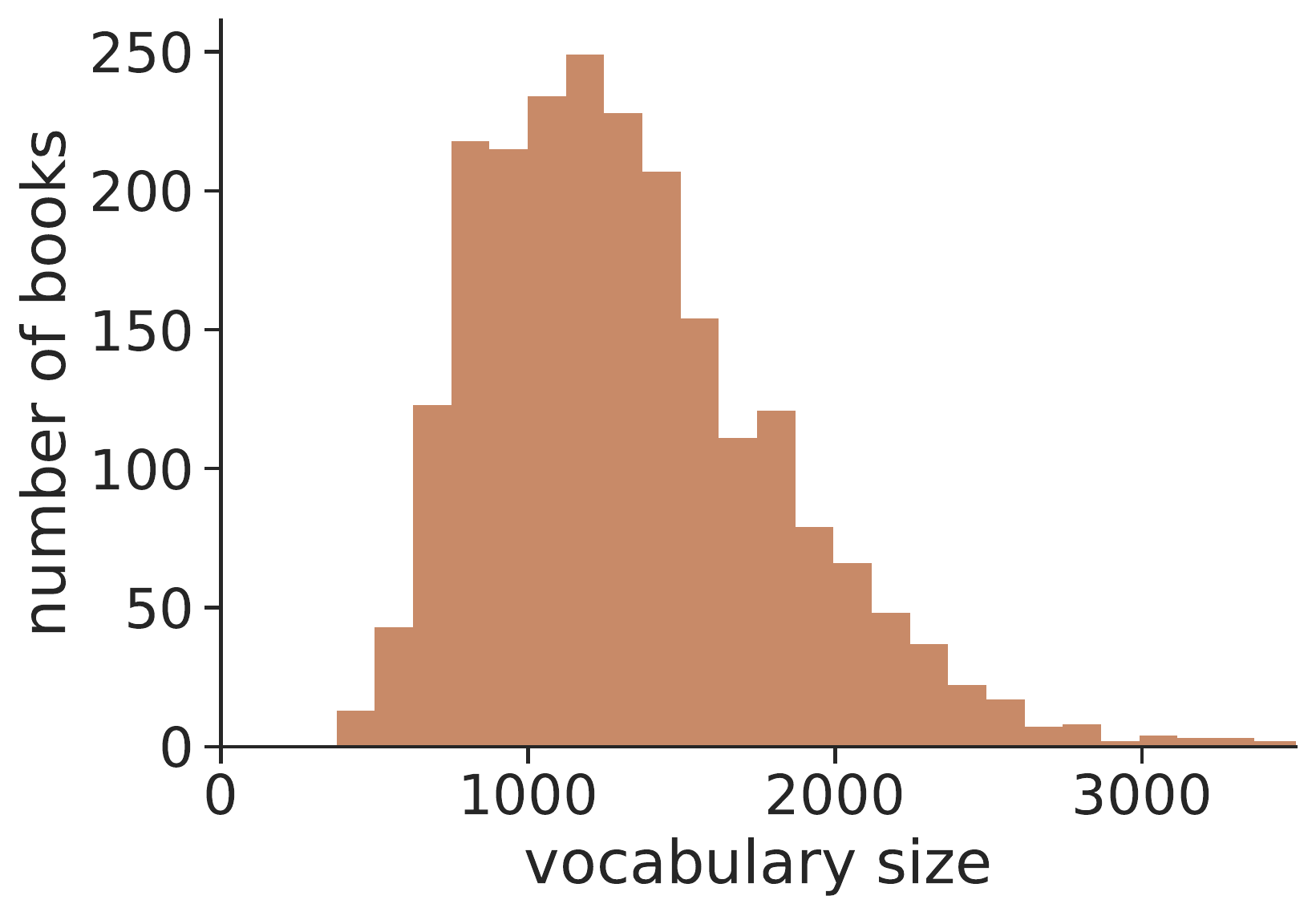}\label{fig:gutenberg_vocabsize}}
\caption{
Compatibility of \B with fiction e-book reading:
Distribution of
(a)~lemma revisitation time and
(b)~vocabulary size
for 2,221 books.
Each book treated as an individual corpus; analysis done for corpus coverage $\alpha=0.9$.
}
\label{fig:gutenberg}
\end{figure}

\xhdr{Corpus generation} 
We use a collection of freely available English\hyp language books from Project Gutenberg \cite{lahiri:2014:SRW}.
After filtering out books with fewer than 25k tokens (about 100 pages), we are left with 2,221 books from 134 authors (median length: 77k tokens; IQR [52k, 113k]).
In our analysis, we fix the text coverage to $\alpha=0.9$ and treat each book as a corpus.
\Figref{fig:gutenberg_vocabsize} contains a distribution over the books' vocabulary sizes at $\alpha=0.9$.

\xhdr{Results}
\Figref{fig:gutenberg_distplot} summarizes the distribution over corpus\hyp wide lemma revisitation times across the 2,221 books (median: 0.55 days; IQR [0.42, 0.70]), showing that in nearly all books, nearly all lemmas tend to reappear within less than a day.
This implies that, whenever the tutor algorithm considers a word worthy of revision, it is likely to become a translation candidate for \B that very day.
This further corroborates the compatibility of \B with everyday information diets.

\section{Discussion}
\label{sec:Discussion}

\xhdr{Summary of results}
While we considered \B to be a potentially effective learning paradigm from the onset, our expectation prior to the user study was that classic, memorization\hyp based vocabulary learning would vastly outperform \B, due to the conscious and focused effort inherent in dedicated learning.

Surprisingly, though, \B led to retention rates competitive with memorization\hyp based vocabulary learning via tables (\Secref{sec:Results RQ1}), both in the short term (where \B surpassed the tables) and in the long term (where the two conditions were indistinguishable).
While Broccoli still incurs a slowdown in reading speed (\Figref{fig:reading_speed_comparison}), the cognitive work causing this slowdown---indicative of a learning process---seems to require substantially less conscious cognitive effort than table-based learning, as indicated by the pronounced lack of consciously used mnemonic strategies (\Figref{fig:strategy_use}), which were, on the contrary, heavily employed in table-based learning.

\xhdrNoPeriod{The surprising effectiveness of Broccoli}
might be due to the fact that incidental learning from context is a very natural mode of learning \cite{nagy1985learning}.
While contextual vocabulary learning is natural and effective, until now it has lacked optimized repetition intervals.
Conversely, while flashcard\hyp based vocabulary learning is able to freely use sophisticated models to schedule words, it lacks the natural mode of learning from context.
Combining context and repetition has been shown to help learning in general \cite{horst2013context}, and we believe that \B's effectiveness stems from applying this approach to vocabulary learning.
Also, \B minimizes the conscious cognitive load by piggybacking on the user's natural Web browsing or e-reading, activities in which they would engage regardless.


\xhdr{Role of language modeling}
We observed the somewhat surprising result that replacing words in contexts for which the language model (\Secref{sec:Context-based word scoring}) predicted a high guessability were \textit{not} retained more easily than words translated in random contexts (\Secref{sec:Results RQ4}).
In our attempts to understand this observation, we hypothesized that the effect of easy contexts might be overpowered by the effect of explicit teaching. To test this hypothesis, we repeated the same analysis as in \Secref{sec:Results RQ4} for the subset of words whose explicit translation was never seen by the respective user (i.e., words on which the user never clicked), but the results remained qualitatively identical, thus refuting our first tentative explanation.

In the initial phase of the project, when exploring how to implement context-based word scoring (\Secref{sec:Context-based word scoring}), we conducted a regression analysis in which the context-based human guessability of hidden words (for which we had collected a small set of ground-truth data) was modeled as a function of features of the hidden word itself and its context.
Inspection of the fitted coefficients revealed that by far the most powerful predictors were the hidden word's frequency in Wikipedia and in the present text, followed by its context-based probability according to the language model.
Since all words used for learning in the user study were by construction frequent in Wikipedia and the input text (\cf\ \Secref{sec:Experimental design}), all words selected for translation in the user study could be expected to have a high guessability by default, such that the marginal contribution of the third most powerful feature---the language model's output---was only minor, which may in turn explain the ineffectiveness of the language model in our particular user study.

If the impact of the language model were to remain low even in real application settings, this would in fact have positive consequences:
it would mean that even a computationally inexpensive, bare-bones implementation of Broccoli, along with a rudimentary replacement strategy, already confers a significant portion of the benefits. This could justify a deployment without a neural\hyp network\hyp based language model, thus allowing Broccoli to work on low-power devices such as cellphones and e-readers with little or no connectivity to cloud computing resources.

In long-form reading such as e-books, one may also leverage strong signals in addition to language models when picking words to translate: the predictability of the reading trajectory enables look-ahead strategies that prepare the reader in a targeted way for the words that will become important in her reading down the line.

\xhdr{Limitations of the user study}
The evaluation of vocabulary retention only tested the direction of Finnish to English, corresponding to a passive knowledge of vocabulary, e.g., being able to understand the foreign language without necessarily being able to speak it.
While this is well aligned with \B's philosophy of learning vocabulary via passive exposure, future work should also evaluate retention in the opposite direction.

The user study was aimed at measuring the efficiency of \B and making it comparable to traditional methods of vocabulary learning.
This has led to the design of our survey with an identical number of words per condition (pre-table, \B, post-table; \cf \Secref{sec:Experimental design}).
A fixed number of words, together with constraints on the duration of the survey (we did not want participants to lose motivation before reaching the post-table) have forced us to condense the interactions with \B.
The text presented during the \B section of the study was selected to allow for a high annotation density, with passive encounters of all words, in a limited time frame.
Normally, when operating on a real-world information diet, we would likely annotate fewer words.

In this light, our study is not able to capture a realistic slowdown induced by \B.
Moreover, participants in our study knew that they would be tested on the vocabulary, which presumably gave them a motivation to focus on translated words and contributed to a higher slowdown recorded in the survey.
At the same time, the results from our study emphasize that learning purely from context, without mnemonic devices and even without ever clicking on a word to reveal its translation, is an effective strategy when using \B. Over time, users might learn to rely more on this passive strategy, which, together with the novelty aspect of \B wearing off, might lead to a reduced slowdown.

In the user study, the tutor algorithm was not used when selecting word occurrences to translate, since spaced repetition works on much longer time scales than our study, which lasted only about an hour.
The effectiveness of spaced repetition has been well established \cite{dempster1989spacing,melton1970situation,godwin2010emerging, ausubel1965effect, wozniak2005two, wozniakTwoComponentsLongterm1995}, and we are confident that \B would profit from the additional influence of a tutor algorithm.
Nevertheless this is an effect that cannot be quantified by the present study.
Future work should therefore conduct long-term, longitudinal studies of the behavior of real-world \B users (beyond students in a lab setting) and of the word retention rates they achieve.

A further minor limitation stems from the fact that the current prototype and the study presented here operate on the level of individual words.
It would, however, be straightforward to extend the method to multi-word collocations and common expressions (\eg, \textit{for instance}, \textit{table tennis}, \textit{birds of a feather}).

\xhdr{Extending the \B paradigm}
Broccoli's design could be evolved in two opposite directions, both toward less and toward more user interaction.
In the direction of \textbf{less interaction,} one could remove the ability to click a word to reveal its translation altogether.
A lack of such a ``temptation'', while occasionally inconvenient, could over time condition the user to ignore the Broccoli aspect of her reading diet to an ever higher degree.
This would be even more aligned with Broccoli's core philosophy, without in general detracting from the learning efficiency, as indicated by the result that learning is effective even without clicking (\Figref{fig:clicking}c--d).

The only user interaction possible in the current prototype and study is clicking on a word to reveal its translation.
In the direction of \textbf{more interaction,} the user interface could be extended to allow the user to provide additional feedback, such as whether a translated word was understandable or not.
Alternatively, one might embed small quizzes into the interface, \eg, asking the user to guess the meaning of a translated word.
Such feedback and gamification elements could keep the user engaged and provide data that could lead to better models of word guessability (language model, \Secref{sec:Context-based word scoring}) and memory (tutor algorithm, \Secref{sec:Word scheduling}).
By keeping such interactions entirely optional, the core idea of \B---learning vocabulary through passive exposure---would remain untainted. 

A long-term, more aspirational goal of a Broccoli-derived approach would be an adjustable annotation density that would very gradually transform the text until culminating in a total conversion from the original language to the target language.
At some point during this process, the translation direction would be flipped: \eg, rather than an originally English corpus being sprinkled with an increasing number of Finnish words, the corpus would at some point become a preeminently Finnish text with a decreasing number of English words. This would also convey an increasing amount of grammatical information, as sequences of adjacent translated words start accumulating.

Whether \B can really be pushed to such a radical level is of course far from certain, but we hope that this paper opens the door to further explorations of creative solutions for sprinkling learning---of vocabulary and beyond---into everyday activities, guiding us systematically to ``learn without intending to learn''.

   

\begin{acks}
We thank Pierre Dillenbourg for valuable feedback on the study design.
This work was supported in part by Microsoft, Google, Facebook, and the Swiss National Science Foundation.
\end{acks}

\bibliographystyle{ACM-Reference-Format}

\balance

\bibliography{references}

\end{document}
\endinput